\documentclass[twocolumn]{aastex63}
\usepackage{color}
\usepackage{comment}
\usepackage{import}
\usepackage{multirow}
\usepackage{CMNDSGen}
\usepackage{CMNDSManual}
\usepackage{CMNDSExtra}
\usepackage{CMNDSZ}
\usepackage{CMNDSZSpecZPhotResub}
\usepackage{CMNDSLBolAGN}
\usepackage{CMNDSEBV}
\usepackage{CMNDSCG}
\usepackage{CMNDSMStar}
\usepackage{CMNDSSFR}
\usepackage{CMNDSCIGALEvSDSS}
\usepackage{CMNDSAssef}
\usepackage{CMNDSScat}
\usepackage{CMNDSZAHatSSFRMstarOffset}
\usepackage{CMNDSCont}
\usepackage{CMNDSFRMstar}
\usepackage{CMNDSLBolZ}
\usepackage{CMNDSLBolAGNFEddEBV}

\shorttitle{WISE AGN Galaxy Catalog}
\shortauthors{Barrows et al.}

\begin{document}

\accepted{for publication in ApJ}

\title{A Catalog of Host Galaxies for \wise-Selected AGN: Connecting Host Properties with Nuclear Activity and Identifying Contaminants}

\author[0000-0002-6212-7328]{R. Scott Barrows}
\affiliation{Department of Astrophysical and Planetary Sciences, University of Colorado Boulder, Boulder, CO 80309, USA}

\author{Julia M. Comerford}
\affiliation{Department of Astrophysical and Planetary Sciences, University of Colorado Boulder, Boulder, CO 80309, USA}

\author[0000-0003-2686-9241]{Daniel Stern}
\affiliation{Jet Propulsion Laboratory, California Institute of Technology, 4800 Oak Grove Drive, Pasadena, CA 91109, USA}

\author[0000-0002-9508-3667]{Roberto J. Assef}
\affiliation{N\'ucleo de Astronom\'ia de la Facultad de Ingenier\'ia y Ciencias, Universidad Diego Portales, Av. Ej\'ercito Libertador 441,
Santiago, Chile}

\correspondingauthor{R. Scott Barrows}
\email{Robert.Barrows@Colorado.edu}

\begin{abstract}
\noindent We present a catalog of physical properties for galaxies hosting active galactic nuclei (AGN) detected by the \wisetitle~(\wise). By fitting broadband spectral energy distributions of sources in the \WAC~\citep{Assef:2018} with empirical galaxy and AGN templates, we derive photometric redshifts, AGN bolometric luminosities, measures of AGN obscuration, host galaxy stellar masses, and host galaxy star formation rates (\SFRs) for \FullSZ~\wise~AGN. The wide-area nature of this catalog significantly augments the known number of obscured AGN out to redshifts of z\,$\sim$\,3 and will be useful for studies focused on AGN or their host galaxy physical properties. We first show that the most likely non-AGN contaminants are galaxies at redshifts of z\,$=$\,0.2\,$-$\,0.3, with relatively blue \wiseonetwo~colors, and with high specific \SFRs~for which the dust continuum emission is elevated in the \wisetwo~filter. Toward increasingly lower redshifts, \wise~AGN host galaxies have systematically lower specific \SFRs, relative to those of normal star forming galaxies, likely due to decreased cold gas fractions and the time delay between global star formation and AGN triggering. Finally, \wise~AGN obscuration is not strongly correlated with AGN bolometric luminosity but shows a significant negative correlation with Eddington ratio. This result is consistent with a version of the `receding torus' model in which the obscuring material is located within the supermassive black hole gravitational sphere of influence and the dust inner radius increases due to radiation pressure.
\end{abstract}


\section{Introduction}
\label{sec:intro}

Correlations between supermassive black holes (SMBHs) and observable properties of their host galaxies \citep[e.g.][]{Gebhardt00,Ferrarese2000,Marconi:Hunt:2003,Bentz:2009c,Gultekin:2009} suggest that evolution of the two may be driven by common mechanisms \citep[e.g.][]{Hernquist:1989,DiMatteo:2005,Springel:2005,Hopkins2008}. However, the number density of AGN, particularly at low luminosities, is dominated by those with significantly attenuated emission \citep[e.g.][]{Merloni:2014,Buchner:2015,Ricci:2017b}, and therefore a significant fraction of SMBH growth may occur in obscured phases (see \citealt{Hickox:2018} for a recent review). The obscuration can be from dust grains absorbing or scattering photons at infra-red (IR) to ultra-violet (UV) wavelengths \citep{Draine:2003a} or by photoelectric absorption and Compton scattering of X-ray photons \citep[see][and references therein]{Comastri:2004}. While much of the obscuring material may reach compact scales of $<$\,1\,pc \citep[e.g.][]{Suganuma:2006,Lopez-Gonzaga:2016} and is often pictured in a torus structure \citep[e.g.][]{Antonucci:1993,Urry:1995,Netzer:2015}, gas and dust throughout the entire galaxy can also contribute to the line-of-sight obscuration. Obscured AGN may therefore preferentially reside in galaxies with large gas fractions that also contribute to star formation and evolution of the host galaxy stellar populations. 

Indeed, the population of nearby ultra-luminous IR galaxies (ULIRGs) are known to have high star formation rates (\SFRs), dust content, and often buried AGN \citep[e.g.][]{Sanders:1988}, and obscured AGN fractions are positively correlated with far-IR (FIR) luminosities that indicate star formation \citep[e.g.][]{Kim:1998,Page:2004, Chen:2015}. Moreover, enhanced star formation is observed in luminous AGN \citep[e.g.][]{Hutchings:1992,Bahcall:1997,Canalizo:2001}, and numerous studies find that AGN bolometric luminosities correlate with host galaxy \SFRs~\citep[e.g.][]{Boyle:1998,Mullaney:2012,Harris:2016,Lanzuisi:2017,Stemo:2020}. However, global star formation and SMBH accretion occur on vastly different physical scales that may introduce temporal delays between the peak of star formation and the most efficient phases of SMBH growth \citep[e.g.][]{Schawinski:2009,Wild:2010,Hopkins:2012a,Barrows:2017b}. 
 
Larger masses of obscuring gas and dust may also contribute to the reservoir of material for SMBH accretion, leading to a positive connection between AGN obscuration and luminosity that is observed in powerful and dust-reddened AGN \citep[e.g.][]{Assef:2015,Tsai:2015,Glikman:2015}. On the other hand, radiation from the AGN can also reduce the level of obscuration through sublimation of dust grains or through radiation pressure \citep[e.g.][]{Lawrence:1991,Simpson:2005,Treister:2008,Assef:2013,Toba:2014,Ricci:2017c}, resulting in the observability of luminous and unobscured AGN as predicted in evolutionary models \citep[e.g.][]{Hopkins05,Younger:2008,Somerville:2008}. 

To study how properties of the obscuring medium are linked with AGN and their host galaxies, observing large populations of obscured AGN is necessary. X-ray photons are often used for surveying obscured AGN \citep[e.g.][]{Civano:2016,Marchesi:2016} since they are less attenuated than optical and UV photons, and they directly trace SMBH accretion disk emission that has been inverse Compton-scattered. Since X-rays can still be scattered and absorbed by large columns of gas, deep X-ray observations are necessary to detect heavily buried AGN. However, a significant fraction of accretion disk emission absorbed by dust is re-emitted at IR wavelengths. This IR emission is therefore correlated with the AGN intrinsic luminosity, as demonstrated by IR bolometric corrections \citep[e.g.][]{Elvis:1994,Richards2006,Hopkins2007,Runnoe:2012} and the strong connection between IR and X-ray emission \citep[e.g.][]{Lutz:2004,Fiore:2009,Lanzuisi:2009,Gandhi:2009,Mateos:2015,Stern:2015}, while being relatively insensitive to further attenuation. Since IR emission is produced by all AGN, it is a powerful tool for finding large numbers of accreting SMBHs, regardless of obscuration. 

Several AGN selection diagnostics (calibrated using `truth' samples of AGN based on X-ray and optical emission line properties) have been developed using mid-IR (MIR) colors from the Infrared Array Camera (IRAC) on the \spitzertitle~\citep{Fazio2004,Lacy:2004,Stern2005,Lacy:2007,Donley:2012}. These diagnostics fundamentally rely on colors to distinguish between the red power-law spectral slopes of AGN and the blackbody spectrum produced by the host galaxy at rest-frame wavelengths of $\lambda$\,$\sim$\,1.6\,$-$\,6\,\micron. More recently, the \wisetitle~\citep[\wise;][]{Wright:2010} has surveyed the entire sky in four bands - \wiseone~(\lameff$\,=3.4\,$\micron), \wisetwo~(\lameff$\,=4.6\,$\micron), \wisethree~(\lameff$\,=12.1\,$\micron), and \wisefour~(\lameff$\,\sim22\,$\micron) - and has enabled development of several MIR AGN selection diagnostics that can be applied to millions of sources \citep{Jarrett:2011,Stern:2012,Mateos:2012,Assef:2013,Secrest:2015}. Among the \wise~bands, this selection is most efficiently accomplished with the \wiseonetwo~color that characterizes the MIR slope for galaxies below $z$$\,\sim\,$1.

Since these MIR diagnostics are based on spectral slopes, they can potentially be mimicked by inactive early-type galaxies at $z$\,$\sim$\,1\,$-$\,2 for which the rising stellar continuum flux at $\lambda$\,$<$\,1.6\,\micron~falls within the \wiseone~and \wisetwo~bands. Moreover, galaxies with significant quantities of dust heated through star formation and to high temperatures can potentially achieve the color requirements of MIR AGN selections (see \citealp{Padovani:2017} for a summary). However, the significant advantages over AGN selection at other wavelengths (comparable sensitivity to both obscured and unobscured AGN) and uniform sensitivity across the entire sky make \wise~ideal for studying the conditions of AGN fueling.

The \WAC~\citep{Assef:2018}, built using a \wiseonetwo~cut with a \wisetwo~magnitude dependence, presents the largest catalog of IR-selected AGN and provides an all-sky catalog of millions of obscured and unobscured AGN candidates. The next step is to determine redshifts and measure physical properties of the AGN and their host galaxies in a uniform and systematic manner. These properties will enable studies of how SMBH growth is connected with host galaxy evolution and how this connection evolves with redshift. Furthermore, to examine these connections under both obscured and unobscured conditions, estimates of obscuration based on multi-wavelength observations are necessary since those based on MIR photometry alone are uncertain \citep[e.g.][]{Hickox:2007,Netzer:2015}. These properties are also necessary to identify populations of potential contaminants, and the redshifts at which they enter the sample, for interpretation of future results. Therefore, in this paper we describe the \WAGC: a catalog of physical properties associated with \wise~AGN and their host galaxies derived from broadband spectral energy distributions (SEDs). We intend for this catalog to be a useful resource for large population studies of \wise~AGN and their host galaxies, and for identification of potential contaminants.

This paper is structured as follows: in Section \ref{sec:catalog} we describe the steps taken to create the catalog, in Section \ref{sec:contamination} we identify the populations of likely contaminants, in Section \ref{sec:Evol} we use the catalog to examine the correlation between star formation and stellar mass for \wise~AGN host galaxies, in Section \ref{sec:SMBHGrowth} we examine the connection between SMBH growth and AGN obscuration, and in Section \ref{sec:conc} we present our conclusions. Throughout we assume a flat cosmology defined by the nine-year Wilkinson Microwave Anisotropy Probe observations \citep{Hinshaw:2013}: \HNaught$\,=\,$\HNaughtValue~\uHNaught~and \OmegaM$\,=\,$\OmegaMValue.

\begin{deluxetable}{ccccccc}
\tabletypesize{\footnotesize}
\tablecolumns{7}
\tablecaption{Summary of Photometry.}
\tablehead{
\colhead{Survey} &
\colhead{Filt.} &
\colhead{$\lambda_{\mathrm{Eff}}$} &
\colhead{FWHM} &
\colhead{Det. Lim.} &
\colhead{$N$} \\
\colhead{($-$)} &
\colhead{($-$)} &
\colhead{(\micron)} &
\colhead{(\micron)} &
\colhead{(Mag)} &
\colhead{($-$)} \\
\colhead{1} &
\colhead{2} &
\colhead{3} &
\colhead{4} &
\colhead{5} &
\colhead{6}
}
\startdata
\multirow{2}{*}{\galex} & \fuv & 0.1516 & 0.02 & 24.2\tablenotemark{a} & 174,186 \\
~ & \nuv & 0.2267 & 0.08 & 24.5\tablenotemark{a} & 583,906 \\ \hline
\multirow{5}{*}{\sdss} & \uband & 0.3543 & 0.06 & 23.5 & 382,619 \\
~ & \gband & 0.4770 & 0.13 & 23.8 & 423,840 \\
~ & \rband & 0.6231 & 0.11 & 23.2 & 434,980 \\
~ & \iband & 0.7625 & 0.12 & 22.6 & 437,347 \\
~ & \zband & 0.9134 & 0.11 & 21.7 & 420,820 \\ \hline
\multirow{2}{*}{\gaia} & \Gbp & 0.532 & 0.31 & 21.8 & 570,205 \\
~ & \Grp & 0.797 & 0.25 & 20.8 & 568,582 \\ \hline
\multirow{5}{*}{\pnstrs} & \gband & 0.4810 & 0.12 & 24.7 & 622,365 \\
~ & \rband & 0.6170 & 0.14 & 24.3 & 675,044 \\
~ & \iband & 0.7520 & 0.13 & 23.4 & 686,233 \\
~ & \zband & 0.8660 & 0.10 & 23.1 & 685,341 \\
~ & \yband & 0.9620 & 0.06 & 22.8 & 644,666 \\ \hline
\multirow{3}{*}{\twomass} & \Jband & 1.2 & 0.2 & 18.6 & 299,882 \\
~ & \Hband & 1.7 & 0.2 & 18.6 & 299,506 \\
~ & \Ksband & 2.2 & 0.3 & 18.5 & 296,301 \\ \hline
\multirow{4}{*}{\wise} & \wiseone & 3.4 & 0.8 & 21.5 & 695,273 \\
~ & \wisetwo & 4.6 & 1.1 & 19.9 & 695,273 \\
~ & \wisethree & 12 & 9 & 18.9 & 694,791 \\
~ & \wisefour & 22 & 5 & 16.9 & ~694,989
\enddata
\tablecomments{Column 1: survey name; column 2: filters from the surveys in column 1; column 3: filter effective wavelengths; column 4: filter full-width at half-maximums (FWHMs); column 5: $5\sigma$ survey/filter AB magnitude detection limits; column $6$: number of sources detected by the survey/filter. \tablenotetext{a}{Since the \galex~coverage has non-uniform depth, the \galex~detection limits correspond to the deepest coverage for our sample.}}
\label{tab:phot}
\end{deluxetable}

\section{Creating the Catalog}
\label{sec:catalog}

In this section we describe our procedure for building the \WAGC. The parent sample is derived from the \WAC~\citep{Assef:2018} in which AGN candidate host galaxies were defined as those with a significant fraction of the total SED luminosity from the intrinsic (unobscured) AGN (\LBolAGN) relative to that of the host galaxy (\LBolHost). This fraction \citep[which correlates with the probability that including an AGN component in the SED model yields a statistically significant improvement;][]{Chung:2014} is defined as \ahat$\,=\,$\AHatDef~with the threshold set at \ahat$\,>\,$\AHatThresh. We use the version of the \WAC~that is optimized for reliable AGN selection at the $90\%$ level (\R) and consists of \WISEAGNR~robust \wise~AGN candidates. The criterion used to select AGN in the \R~version is a \wiseonetwo~lower threshold that is a function of \wisetwo. We refer the reader to \citet{Assef:2018} for further details on the \wise~AGN candidate selection function and additional filters applied to develop the final \WAC.

The specific steps we take are as follows: identifying the subset of the \R~\WAC~with uniform multi-wavelength coverage (Section \ref{sec:PhotData}), matching with multi-wavelength detections (Section \ref{sec:FiltMatchPhot}), modeling the SEDs to measure photometric redshifts and to separate the galaxy and AGN contributions (Section \ref{sec:redshifts}), applying a quality filter to generate the final sample (Section \ref{sec:final}), quantifying the photometric redshift accuracy (Section \ref{sec:PhotoZAccuracy}), using the best-fit SED models to compute intrinsic physical properties of the AGN (Section \ref{sec:AGNProps}) and of the host galaxies (Section \ref{sec:GalProps}), and estimating the impact of scattered AGN light (Section \ref{sec:scattered}). In Section \ref{sec:Cat} we describe the data structure of the final catalog.

\subsection{Photometric Datasets}
\label{sec:PhotData}

We use photometric coverage from five additional major astronomical surveys: the \galextitle~(\galex), the \sdsstitle~Data Release 14 (\sdss~DR14), \gaia~DR2, the \pnstrstitle~DR2 (\pnstrs2), and the \twomasstitle~(\twomass) Point Source Catalog. These photometric databases are chosen for two primary reasons: 1) they each have wide-area sky coverage; and 2) when combined with the \wise~photometry, the full filter set covers \fuvLamEffMicron$\,-\,$\wisefourLamEffMicron\,\micron~(most of the rest-frame \LRTRange~range covered by our SED models). We require that each \R~\wise~AGN candidate have coverage at UV (\galex), optical (\gaia~and \sdss~or \pnstrs), NIR (\twomass), and MIR (\wise) wavelengths. If coverage is available from a survey but no detection exists, we incorporate flux upper limits into the SED modeling based on the $5\sigma$ detection limits. The surveys/filters used in our analysis, their effective wavelengths, filter full-width at half-maximums (FWHMs), and detection limits are listed in Table \ref{tab:phot}.

\begin{figure}[t!]
\includegraphics[width=0.48\textwidth]{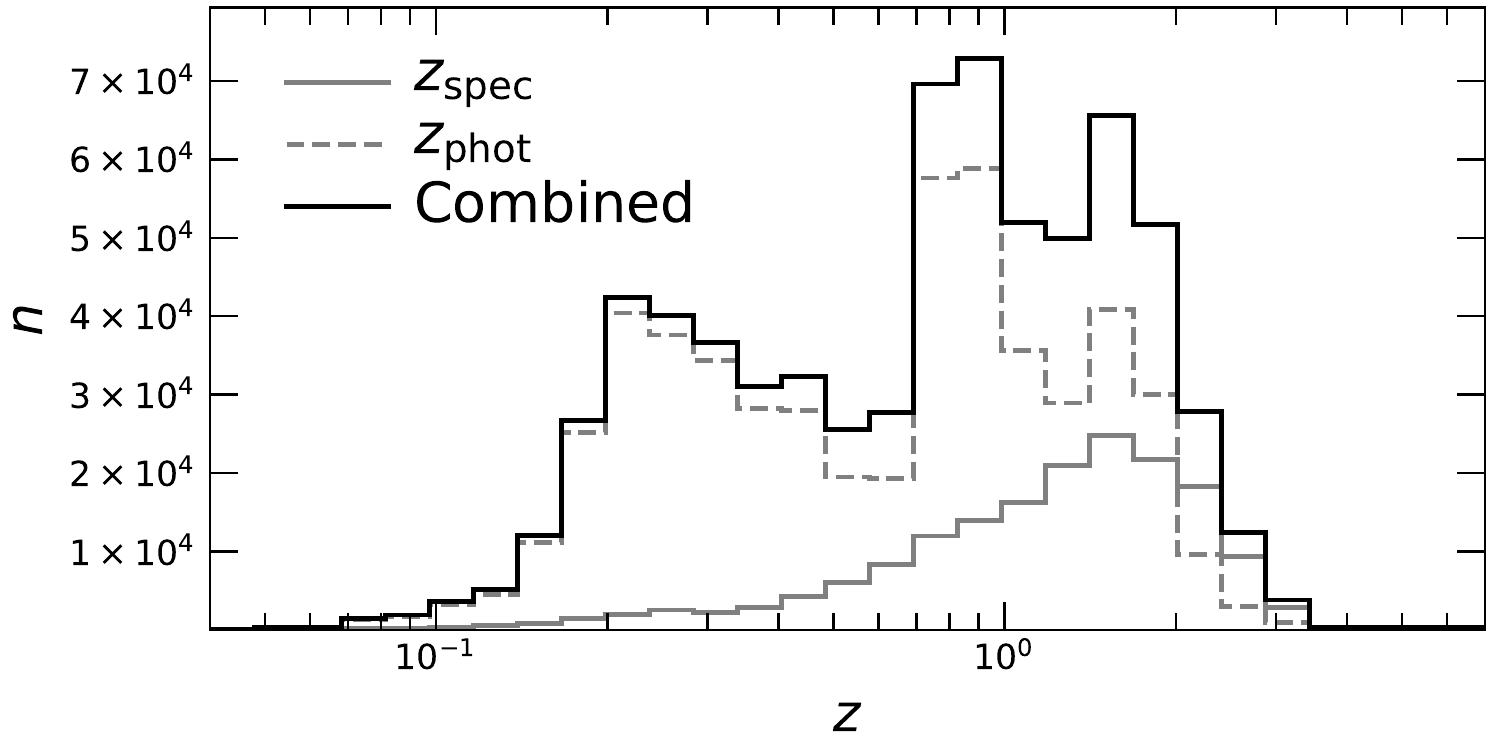}
\caption{\footnotesize{Distribution of redshifts (\z) for the final sample (Section \ref{sec:final}). The subsets with spectroscopic redshifts (\ZSpec) and with only photometric redshifts (\ZPhot) are shown with the solid and dashed gray lines, respectively. The full sample is shown with a solid black line. The spectroscopic subset is biased toward larger redshifts (median of \ZSpec\,$=$\,\SpecZMedian) compared to the subset with only photometric redshifts (median of \ZPhot\,$=$\,\PhotZMedian).}}
\label{fig:Z_HIST}
\end{figure}

\begin{figure}[t!]
\includegraphics[width=0.49\textwidth]{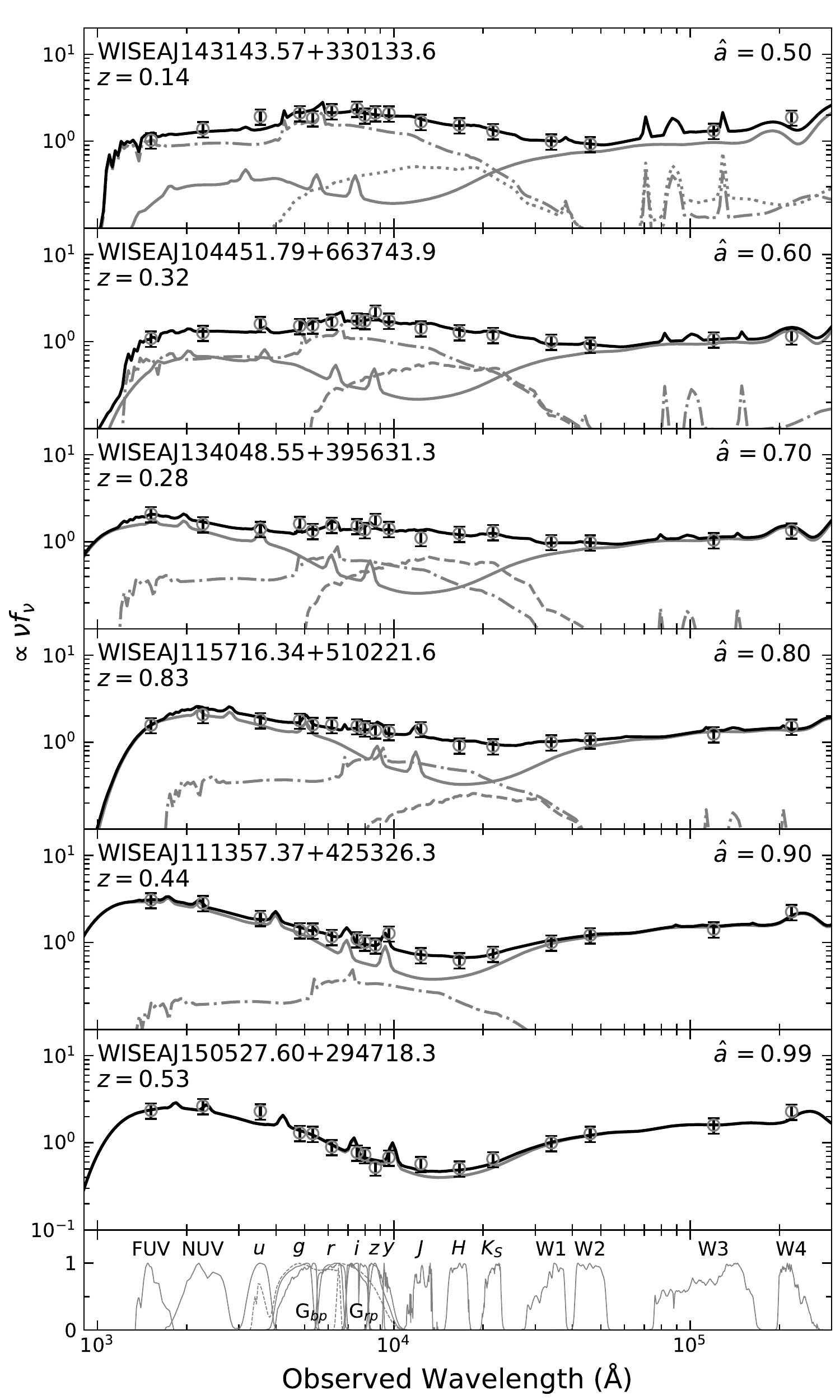}
\caption{\footnotesize{SEDs and best-fit models (normalized at 3.6\,\micron) for \wise~AGN candidates in our final catalog (Section \ref{sec:final}). These six examples are chosen to represent AGN fractions (\ahat) over the range \ahat$\,=0.5-1$ and to have 17 photometric detections (for demonstrating the complete multi-wavelength data set). For consistent comparison, these examples are drawn from the spectroscopic subset. Data points (with $1\sigma$ error bars) are shown as open circles. The AGN, elliptical, Sbc, and Irregular templates are shown as solid, dashed, dotted, and dash-dotted gray lines, respectively, while the best-fit sum is shown as a solid black line. The response curves for each filter (Table \ref{tab:phot}) are shown at the bottom (normalized to the same peak value). Regardless of AGN obscuration level, \wise~fluxes are dominated by the AGN, though the host galaxy does contribute at lower \ahat~values.}}
\label{fig:SEDs}
\end{figure}

\subsection{Matching and Filtering the Photometry}
\label{sec:FiltMatchPhot}

We assign photometric detections from the surveys described in Section \ref{sec:PhotData} to unique \wise~AGN candidates based on a fixed angular separation upper limit as in \citet{Assef:2013,Assef:2018}. Based on photometric redshift accuracy estimates (Section \ref{sec:PhotoZAccuracy}) for 10\%~of the parent sample, we find that angular separation upper limits larger than \MatchThreshAS~correspond to a significant decrease in photometric redshift quality, indicating that the incidence of spurious matches may increase for larger thresholds. Angular separation upper limits lower than \MatchThreshAS~do not result in significantly improved or reduced photometric redshift quality, though the number of matched sources is significantly reduced.  Therefore, to optimize both the number of sources with sufficient data points for SED modeling and their photometric redshift accuracy, we adopt a fixed angular separation upper limit of \MatchThreshAS.

The absolute astrometric accuracy of the \sdss~\citep[$\lesssim\,$\sdssAstrErr;][]{Pier:2003} and \twomass~\citep[$\lesssim\,$\twomassAstrErr;][]{Skrutskie:2006} are both tied to the Tycho reference frame \citep{Hog:2000}, while that of \wise~\citep[$\lesssim\,$\wiseAstrErr;][]{Wright:2010} is tied to \twomass. The absolute astrometry of \galex~\citep[$\sim\,$\galexAstrErr;][]{Morrissey2007} is tied to the U.S. Naval Observatory Catalog \citep{Zacharias:2000}, though it is shown to be consistent with the \sdss~to within $<0\farcs3$. The absolute astrometry of \pnstrs~\citep[$\lesssim\,$\pnstrsAstrErr;][]{Chambers:2016,Magnier:2020} is tied to that of \gaia~DR1 which itself is tied to the Tycho catalog. Finally, the absolute astrometry of \gaia~DR2 \citep[$<\,$1\,mas;][]{Lindegren:2018}, is independent of these catalogs. Since the absolute astrometric accuracies of these surveys are smaller than their typical positional uncertainties, further registration of the datasets is not necessary.

If more than one source from any survey is within the \MatchThreshAS~matching threshold, the \wise~AGN candidate is not passed to the SED fitting process on the justification that multiple sources are blended and that a reliable SED model can not be generated. To mitigate the effects of variability, if detections in multiple epochs are available from a given survey, then photometry from the observations taken closest in time to the corresponding \allwise\footnote[1]{\href{https://wise2.ipac.caltech.edu/docs/release/allwise/}{https://wise2.ipac.caltech.edu/docs/release/allwise/}} observation is used. Furthermore, since the \gband, \rband, \iband, and \zband~filters from \pnstrs~and \sdss~overlap, of those filters we only use detections from the one taken closest in time to the \allwise~observation. An SED is passed to the modeling step (Section \ref{sec:redshifts}) if it contains $\geq\,$\MinPhotNum~photometric data points. We choose this threshold because it corresponds to the minimum number of data points necessary to generate a statistically meaningful model.

\subsection{Photometric Redshifts and SED Models}
\label{sec:redshifts}
 
To compute photometric redshifts (\ZPhot) and model the SEDs of \wise~AGN candidates that pass the photometry requirements outlined in Sections \ref{sec:PhotData} and \ref{sec:FiltMatchPhot}, we use the galaxy and AGN \LRTtitle\footnote[2]{\href{\lrtlink}{\lrtlink}} (\lrt) from \citet{Assef2010}. The \lrt~libraries consist of empirical galaxy templates (elliptical, Sbc spiral, and irregular) plus a \typeI~AGN template to which variable extinction is applied (accounting for nuclear obscuration), and we use them for our SED modeling because they were used to select reliable AGN for the \WAC.

The templates were modeled after the SEDs of galaxies in the AGN and Galaxy Evolution Survey \citep[AGES;][]{Kochanek:2012}, and we refer the reader to \citealt{Assef2010} for more details on their construction. Here we summarize the basic properties: the galaxy templates were adopted from \citet{Coleman:1980} and supplemented with synthesized stellar templates at UV and IR wavelengths \citep[from][]{Bruzual:Charlot:2003} plus MIR SEDs from \citet{Devriendt:1999} that include dust and polycyclic aromatic hydrocarbon (PAH) emission. The AGN template is a combination of power laws plus broad emission lines, modeled on the composite \typeI~quasi-stellar object (QSO) template from \citet{Richards2006}. 

\begin{figure}[t!]
\includegraphics[width=0.48\textwidth]{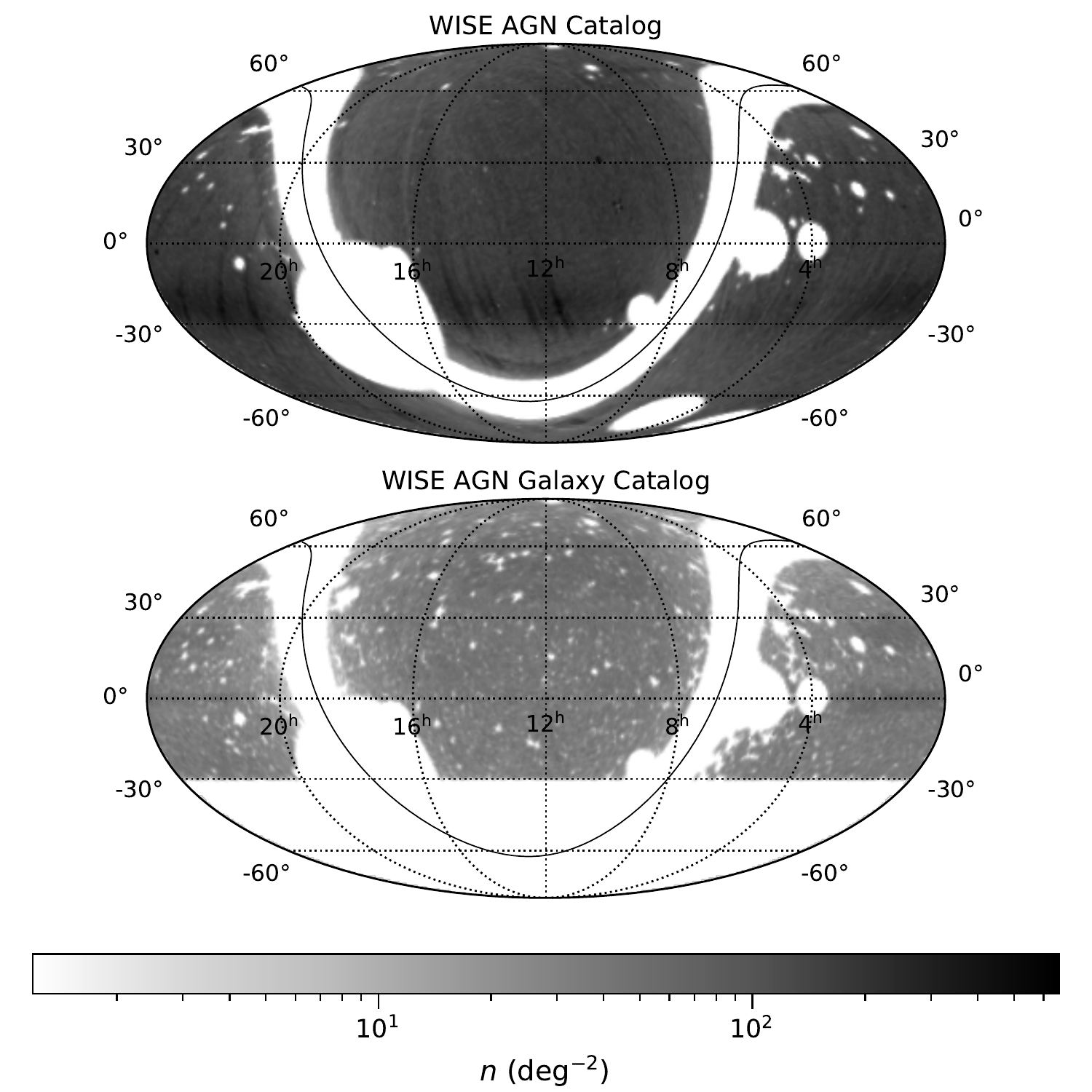}
\caption{\footnotesize{Surface density sky map of sources in the \WAC~(top) and our catalog (bottom). Each panel is in equatorial coordinates and displayed with a Mollweide projection. The solid line indicates the Galactic Plane. The absence of sources below declinations of $\delta$\,$=$\,-30$^{\circ}$ in our catalog is due to the survey coverage requirements (Section \ref{sec:PhotData}).}}
\label{fig:MOC_PLOT}
\end{figure}

The SED modeling procedure we use computes optimal linear combinations of the templates by fitting to the observed photometry and varying six parameters to minimize the \chisqr~statistic: the redshift, normalizations for each of the four templates, and the color excess (\EBV) that controls the wavelength-dependent extinction law applied to the \typeI~AGN component. The extinction law was constructed by \citet{Assef2010} from the \citet{Cardelli:1989} optical-IR function ($\lambda$\,$>$\,3,300\,\AA) and the \citet{Gordon:Clayton:1998} UV function ($\lambda$\,$<$\,3,300\,\AA) and assumes a Milky Way extinction curved defined by \RV$\,=\,$\RvValue~\citep{Schultz:1975}. We apply a standard mean intergalactic medium (IGM) absorption to each galaxy. All photometric fluxes have been corrected for Galactic extinction using the dust maps of \citet{Schlegel98} and \citet{Schlafly:2011}.

To improve the \ZPhot~accuracy, during the fitting we apply a luminosity prior to the galaxy components in the form of a redshift probability function \citep{Assef:2008}: $P(z)\propto e^{-\chi^2(z)/2}\Phi(M)dV_{\mathrm{com}}(z)$ where $dV_{\mathrm{com}}$ is the comoving volume per unit redshift, and $\Phi$ is the Las Campanas Redshift Survey $R-$band luminosity function from \citet{Lin:1996}, defined by a Schechter function \citep{Schechter:1976} with a characteristic magnitude of $M^{\star}=-21.4$ and a slope of $\alpha=-0.7$ \citep{Assef2010,Chung:2014,Carroll:2021}. We allow the models to explore redshifts in the range \ZPhot$\,=\,$0\,$-$\,7 with a grid of step-size $\Delta\,$\ZPhot$\,=\,$\zStep~(significantly smaller than the estimated photometric redshift uncertainties; Section \ref{sec:PhotoZAccuracy}). In no cases do the \ZPhot~solutions converge on the upper bounds (\ZPhot\,$=$\,7).

The wide range of possible dust temperatures, contributions from host galaxy star formation, and silicate emission introduce spectral diversity into AGN SEDs at wavelengths longer than $\sim\,$20\,\micron~\citep[e.g.][]{Mason:2012,Dale:2014,Ly:2018} that is difficult for the \lrt~to model. Therefore, as in \citet{Assef2010,Assef:2013} we remove \wisefour~photometry when solving for \ZPhot~(removal of the \wisefour~filter is accounted for in the requirement of $\ge$\,6 photometric data points outlined in Section \ref{sec:FiltMatchPhot}).

To identify the subset with spectroscopic redshifts (\ZSpec), we crossmatch with the \sdss~DR16 spectroscopic database\footnote[3]{The \sdss~DR16 is chosen since it has the largest catalog of uniformly-measured spectroscopic redshifts that overlap with the \WAC.} using the same \MatchThreshAS~matching radius as in Section \ref{sec:FiltMatchPhot}  
and select sources with primary spectroscopic classifications of either \GAL~or \QSO~\citep[obtained from template fits;][]{Bolton:2012}. If a \ZSpec~value is available, it is adopted as the final redshift (\z). Otherwise, the \ZPhot~value is used.

\begin{figure}[t!]
\includegraphics[width=0.48\textwidth]{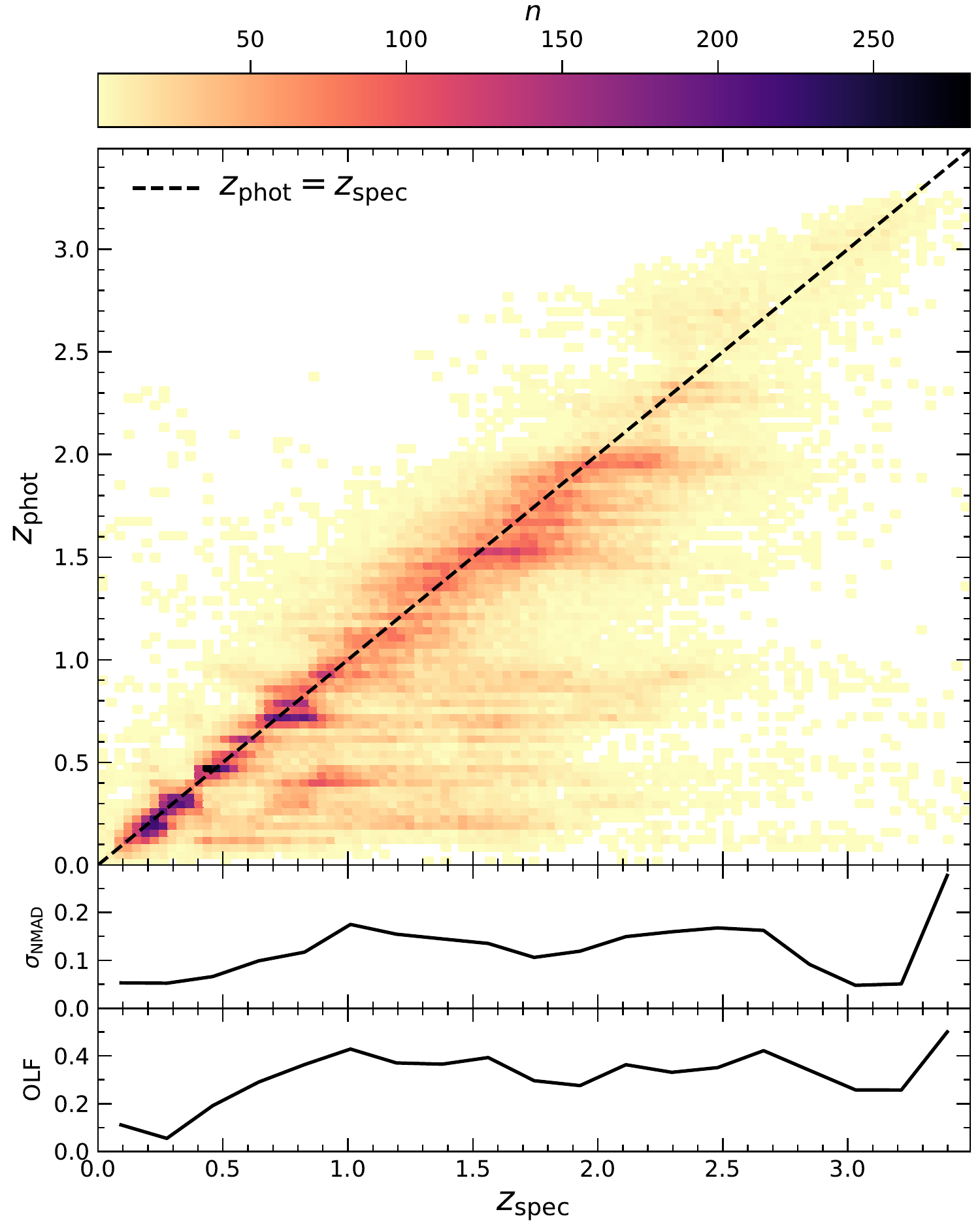}
\caption{\footnotesize{Top: photometric redshift (\ZPhot) as a function of spectroscopic redshift (\ZSpec) for sources in the \sdss~spectroscopic subset of our catalog. The overall \ZPhot~accuracy is \SigNMADSymb$\,=\,$\SigNMAD~with an outlier fraction of \OLFSymb$\,=\,$\OLF. Middle and bottom: \SigNMADSymb~and \OLFSymb, respectively, as a function of \ZSpec. The largest \SigNMADSymb~and \OLFSymb~values are associated with systematically under-estimated photometric redshifts at \ZPhot$\,\lesssim1$.}}
\label{fig:ZSPEC_PHOTOZ_PLOT}
\end{figure}

\begin{figure}[t!]
\includegraphics[width=0.48\textwidth]{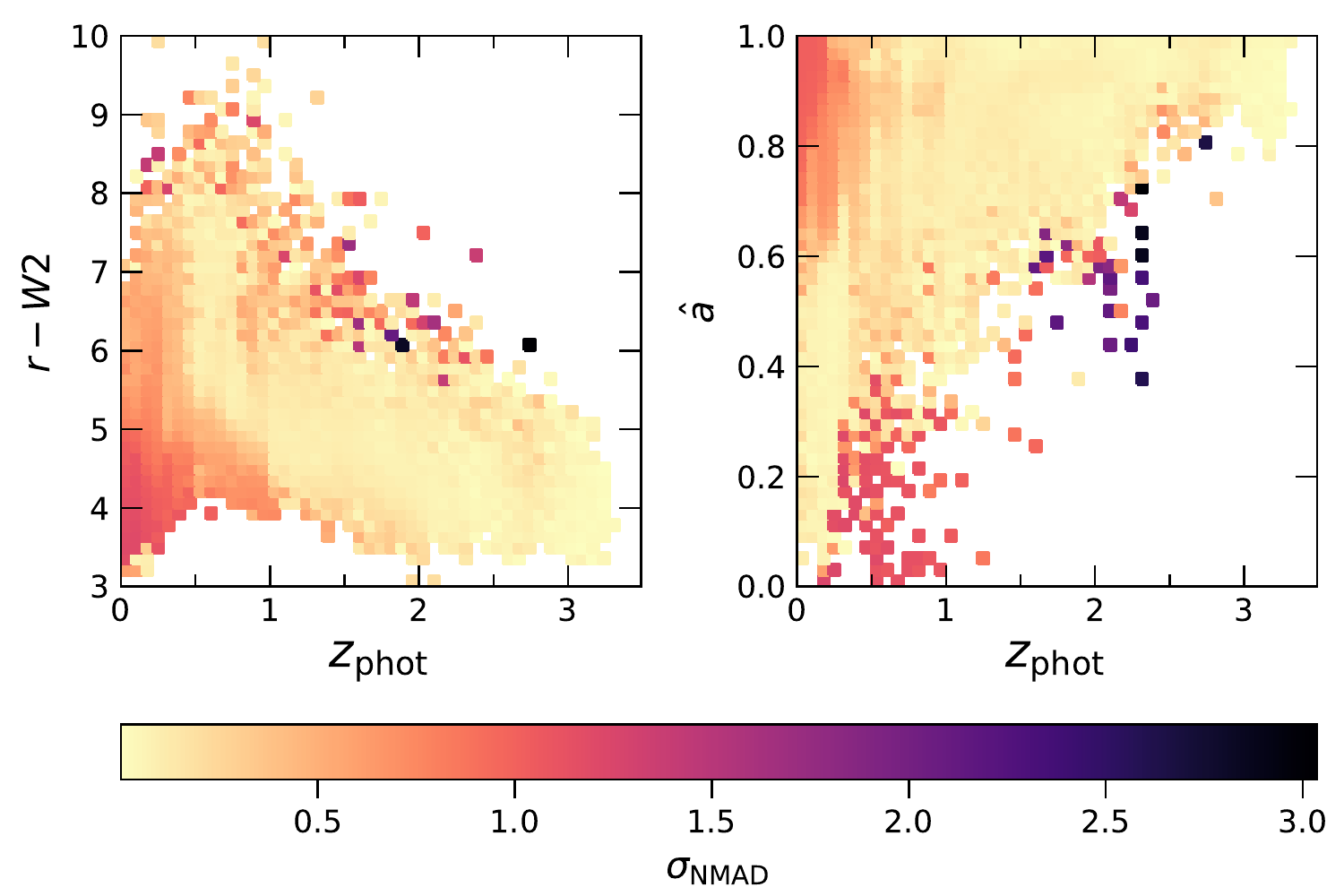}
\caption{\footnotesize{Color (\rwtwo; left) and AGN fraction (\ahat; right) as a function of photometric redshift (\ZPhot). The color-scale represents the \ZPhot~accuracy (\SigNMADSymb). The poorest accuracy occurs at low \ZPhot~values for sources with blue \rwtwo~colors or with large \ahat~values that correspond to the systematic outliers observed in Figure \ref{fig:ZSPEC_PHOTOZ_PLOT}.}}
\label{fig:RW2_AHAT_ZPHOT_PLOT}
\end{figure}

\subsection{Final Sample of SED models}
\label{sec:final}

To obtain the final SED models that separate the galaxy and AGN contributions, we re-fit each SED fixing the redshift at the final value (\z) and varying the remaining parameters (including the IGM absorption). To remove outlying poor SED fits, we require the reduced $\chi^2$ ($\chi_{\mathrm{red}}^2$) to be $\chi_{\mathrm{red}}^2<20$ \citep{Carroll:2021}. Furthermore, we remove the \UPercLumErrCut$\%$ with the largest combined galaxy and AGN component luminosity errors to filter out unreliable decompositions. Imposing these restrictions yields a final sample of \FullSZ~\wise~AGN candidates with SED models. The median $\chi_{\mathrm{red}}^2$ value is 1.52, with a standard deviation of 1.50. The subset with spectroscopic redshifts (\ZSpecSZ~sources) has a median value of \ZSpec\,$=$\,\SpecZMedian, and the subset with only photometric redshifts (\ZPhotSZ~sources) has a median value of \ZPhot\,$=$\,\PhotZMedian~(Figure \ref{fig:Z_HIST}). The spectroscopic subset is dominated by the \QSO~class (\QSOPerc\%) which is likely responsible for the higher redshifts (though note that a fraction of the \ZPhot~values are under-estimated; see Section \ref{sec:PhotoZAccuracy}). A local peak is observed around \z\,$\sim$\,0.25 \citep[also observed in][]{Assef:2013,Assef:2018} that may be due, in part, to contaminants (see Section \ref{sec:contamination}). SEDs with \ahat~values in the range \ahat\,$=$\,0.5\,$-$\,1 are displayed in Figure \ref{fig:SEDs} and show that the \wise~filter fluxes are dominated by the AGN. Figure \ref{fig:MOC_PLOT} shows the surface density sky maps for the parent sample (the \WAC) and for our catalog which is limited to declinations of $\delta$\,$>$\,-30$^{\circ}$ due to the survey coverage requirements (Section \ref{sec:PhotData}). The numbers of sources detected in each survey/filter combination are listed in Table \ref{tab:phot}.

\subsection{Photometric Redshift Accuracy}
\label{sec:PhotoZAccuracy}

We use the subset with \ZSpec~values (Section \ref{sec:redshifts}) to quantify the photometric redshift accuracy (Figure \ref{fig:ZSPEC_PHOTOZ_PLOT}). When parameterized with the normalized median absolute deviation \citep[NMAD;][]{Hoaglin:1983} metric of \SigNMADSymb$\,=\,$\SigNMADDef~where \DeltaZSymb$\,=\,$\DeltaZDef, we find \SigNMADSymb$\,=\,$\SigNMAD. We also find that the outlier fraction (\OLFSymb; defined as the fraction with \DeltaZNormDef$\,>\,$0.15) is \OLFSymb$\,=\,$\OLF. Since our catalog is limited to photometry from broad filters, the \ZPhot~accuracy is worse than for AGN in the Cosmic Evolution Survey \citep[COSMOS;][]{Scoville:2007} field for which medium filters are available \citep[e.g. \SigNMADSymb$\,=0.014$;][]{Salvato:2009}. However, the accuracy is comparable to that obtained from other common template sets that are fit to IR-selected AGN host galaxies and limited to broadband photometry \citep[e.g. \SigNMADSymb$\,=\,$0.095\,$-$\,0.29 and \OLFSymb$\,=\,$\,0.31\,$-$\,0.51;][]{Duncan:2018,Brown:2019}. The systematically under-estimated \ZPhot~values of the outliers are caused by the prior that restricts galaxy masses from being unphysically large \citep{Assef2010,Assef:2013}.

Photometric redshifts of galaxies with observed SEDs dominated by AGN (i.e. hosts of unobscured AGN) are typically less accurate than those of SEDs with detectable features of the galaxy stellar continua \citep[e.g.][]{Brodwin:2006,Rowan-Robinson:2008,Salvato:2009}. Therefore, in Figure \ref{fig:RW2_AHAT_ZPHOT_PLOT} (left) we show how \SigNMADSymb~varies as a function of \ZPhot~and color (\rwtwo; useful as a non-parametric proxy for AGN obscuration). In Figure \ref{fig:RW2_AHAT_ZPHOT_PLOT} (right) we also show how \SigNMADSymb~varies as a function of \ZPhot~and \ahat~since the AGN fraction is shown to be strongly correlated with \ZPhot~accuracy from the \lrt~and is relatively stable against inaccurate \ZPhot~estimates \citep{Assef2010,Assef:2013}. While the poorest accuracy does indeed occur among systems with blue \rwtwo~colors and large \ahat~values, this effect is mainly seen at low \ZPhot~values due to the systematic effects of the prior. Above \ZPhot$\,\sim\,$1 the photometric redshift accuracy is relatively stable, regardless of the \rwtwo~or \ahat~values. In Sections \ref{sec:AGNProps} and \ref{sec:GalProps} we discuss the effects of these systematic offsets on the derived physical properties.

\subsection{AGN Bolometric Luminosities and Obscuration}
\label{sec:AGNProps}

\begin{figure}[t!]
\includegraphics[width=0.49\textwidth]{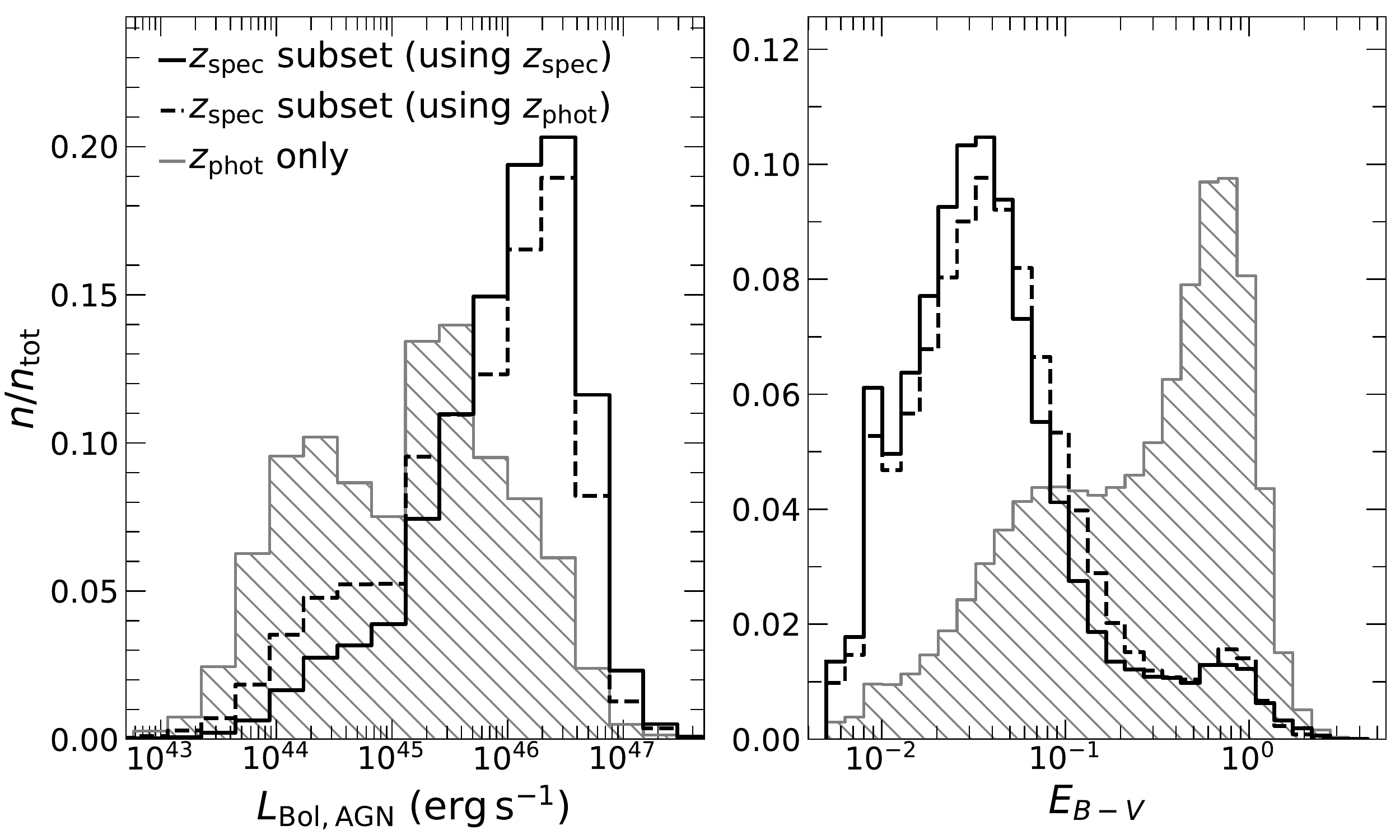}
\caption{\footnotesize{Distribution of \wise~AGN bolometric luminosities (\LBolAGN; left) and extinctions (\EBV; right). The subset with \ZSpec~values is shown in black (solid: using \ZSpec; dashed: using \ZPhot). The subset without \ZSpec~values is shown in gray (hatched). Each sample is normalized to an integrated value of unity. The systematic offsets due to \ZPhot~uncertainties tend toward under-estimated \LBolAGN~values and over-estimated \EBV~values. The subset with \ZSpec~values is dominated by the \QSO~class and has a larger fraction of luminous and unobscured AGN.}}
\label{fig:SEDProps_AGN_HIST}
\end{figure}

AGN bolometric luminosities (\LBolAGN) are computed from the un-extinguished, normalized best-fit AGN templates integrated between the full rest-frame wavelength range of the \lrt~libraries (\LRTRange). AGN extinctions are parameterized by the color excess term (\EBV). The distributions of \LBolAGN~and \EBV~for our catalog are shown in Figure \ref{fig:SEDProps_AGN_HIST}. The \ZPhot~uncertainties introduce a systematic negative offset in the \LBolAGN~values (median offset of \LBolAGNSysOffsetPercMedian\%) that is primarily driven by the population with under-estimated photometric redshifts at \ZPhot$\,\lesssim$\,1 (see Figures \ref{fig:ZSPEC_PHOTOZ_PLOT} and \ref{fig:RW2_AHAT_ZPHOT_PLOT}). On the other hand, the values of \EBV~are relatively stable against \ZPhot~uncertainties (median offset of \EBVSysOffsetPercMedian\%).

\begin{figure*}[t!]
\includegraphics[width=0.98\textwidth]{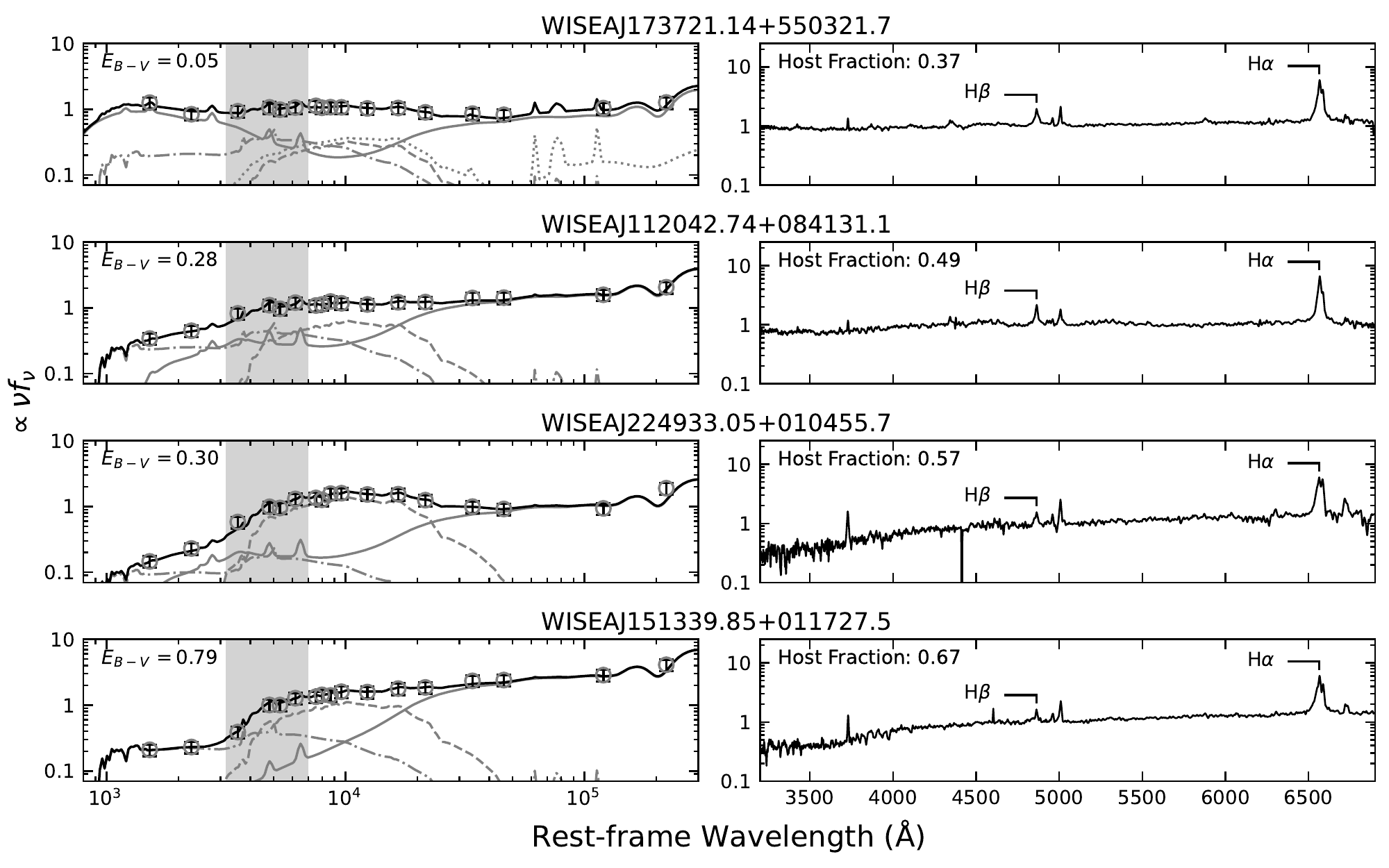}
\caption{\footnotesize{Left: Rest-frame, $k-$corrected SEDs and best-fit models from the \QSO~sample of our catalog (photometry and line styles are the same as in Figure \ref{fig:SEDs}). Right: \sdss~spectra displayed over the rest-frame wavelength range indicated by the shaded region in the left panels. In all panels the flux densities are normalized at rest-frame 5100\,$\mathrm{\AA}$. The examples shown have been limited to $z<0.35$ to include spectral access to both the H$\alpha$ and H$\beta$ emission lines. Ordered by increasing color excess (\EBV) from top to bottom, they represent the 50th, 90th, 95th, and 99th percentiles. The host galaxy fractions (as measured from the \sdss~spectra) increase with \EBV~(as measured from our SED models). While the AGN continua can be subject to significant extinction, they are sufficiently luminous (relative to the host galaxy) for broad emission lines to be detected.}}
\label{fig:ObsQSOs}
\end{figure*}

After correcting for systematic offsets due to \ZPhot~uncertainties, the subset with only photometric redshifts has smaller \LBolAGN~values (median of \LBolAGN\,$=$\,\LBolAGNMedianCorrPhot\,\uLum) and larger \EBV~values (median of \EBV\,$=$\,\EBVMedianCorrPhot) compared to the spectroscopic subset (median values of \LBolAGN\,$=$\,\LBolAGNMedianSpec\,\uLum~and \EBV\,$=$\,\EBVMedianSpec). These differences reflect that the subset with \ZSpec~values is dominated by the \QSO~class and hence toward more luminous and unobscured AGN. The photometric redshift sample therefore significantly augments the number of luminous and obscured AGN.

While sources in the spectroscopic \QSO~class have broad emission lines by definition, \QSOHiEBVPerc$\%$ of them have obscuration levels larger than the typically assumed value of \NH$\,=10^{22}\,$\uNH~(or \EBV$\,\approx\,$0.2, using the gas-to-dust conversion of \citealp{Maiolino:2001}) that broadly separates \typeI~and \typeII~AGN \citep[e.g.][]{Ueda:2003,Burtsche:2016,Schnorr-Muller:2016}. However, large intrinsic ratios of AGN to host galaxy luminosity can result in spectroscopic detections of broad emission lines despite large nuclear obscuration \citep[see][and references therein]{Hickox:2018}. The obscuration will be manifested in spectra as stronger galaxy stellar continuum features at rest-frame optical and UV wavelengths. Indeed, as demonstrated in Figure \ref{fig:ObsQSOs}, the subset of QSOs with large nuclear obscuration as measured from our SED models (\EBV$\,\ge0.2$), have larger host galaxy fractions (by an average of \ObscQSOHostPercIncrease$\%$) as measured from the optical \sdss~spectra \citep{Rakshit:2020}. 

Figure \ref{fig:ObsQSOs} also shows that broad emission lines are only weakly present (or not present at all) in our summed SED models, and this difference likely reflects the limitations of our AGN template. The SED models may incorrectly fit the UV flux with a star-forming component that artificially lowers the AGN contribution by adding nuclear extinction. However, this effect is countered by the inherent prior in our SED modeling against highly reddened AGN models that improve the fit at MIR wavelengths without contributing to the flux at shorter wavelengths (see \citealt{Assef2010} for more details on this prior). Therefore, the true AGN components of these systems may instead have stronger optical broad emission lines than assumed in our models.

\subsection{Host Galaxy Physical Properties}
\label{sec:GalProps}

In this section we discuss the host galaxy stellar masses (\Mstar) and star formation rates (\SFRs) for our catalog of \wise~AGN candidate host galaxies. In Section \ref{sec:SFR} we describe our procedure for estimating \Mstar~and \SFR, and in Section \ref{sec:comp} we compare our estimates with those from independent methods.

\subsubsection{Stellar Mass and Star Formation Rate Estimates}
\label{sec:SFR}

We first subtract the AGN contribution (determined from the \lrt~models; Section \ref{sec:final}) from each photometric detection and compute physical models for the host galaxies using the Code Investigating GALaxy Emission \citep[\cigale;][]{Noll:2009,Boquien:2019} with a delayed star formation history, a Salpeter initial mass function \citep{Salpeter:1955}, and the stellar population libraries of \citet{Bruzual:Charlot:2003}. To use the results from these models we require that $L_{\rm{Bol,Host}}$ is measured at $>\,$\GalSumSigFacThresh$\sigma$ significance (\GalDetSZ~\wise~AGN candidates) since galaxy properties can not otherwise be reliably determined.

\begin{figure}[t!]
\includegraphics[width=0.49\textwidth]{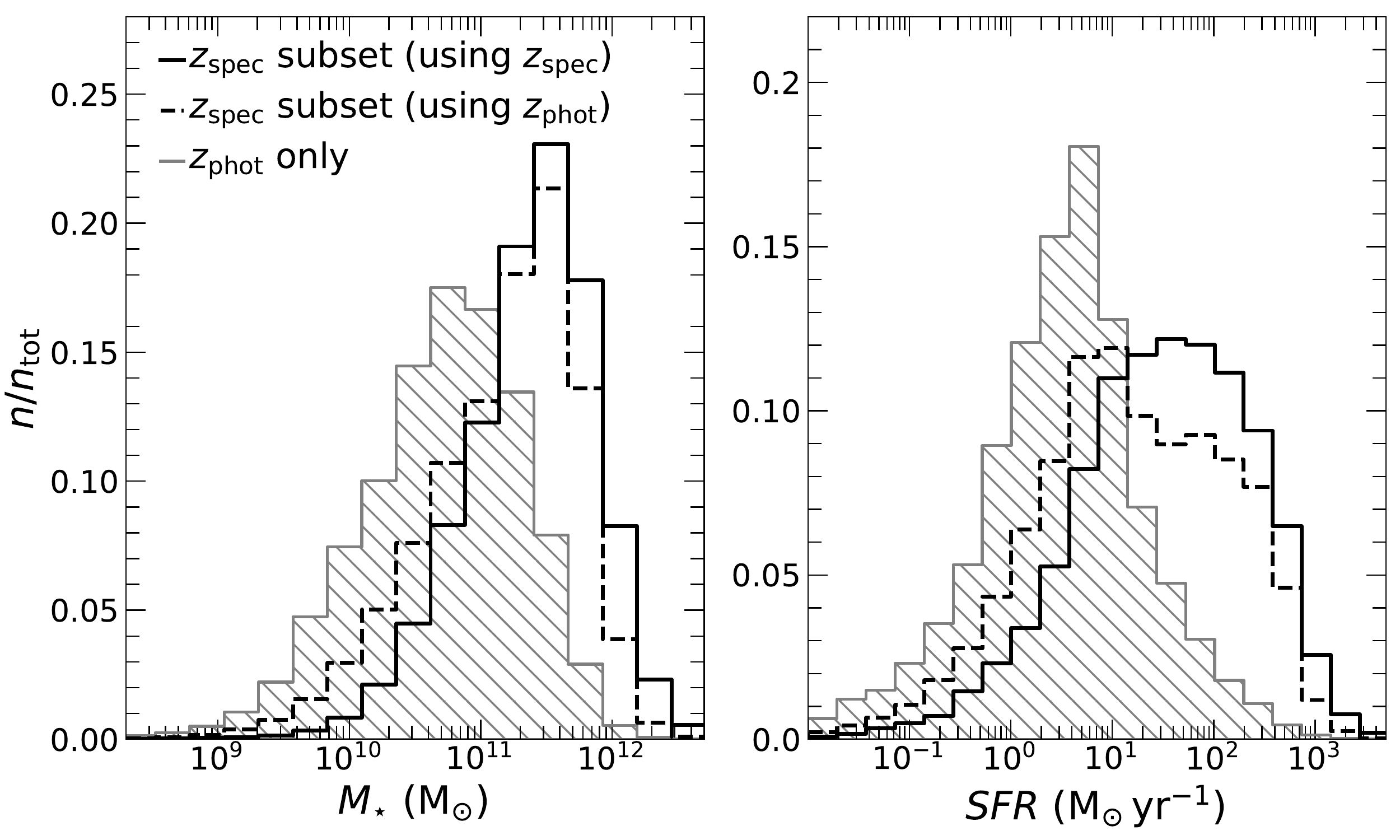}
\caption{\footnotesize{Distribution of \wise~AGN  stellar masses (\Mstar; left) and star formation rates (\SFRs; right). The subset with \ZSpec~values is shown in black (solid: using \ZSpec; dashed: using \ZPhot). The subset without \ZSpec~values is shown in gray (hatched). Each sample is normalized to an integrated value of unity. The systematic offsets due to \ZPhot~uncertainties tend toward under-estimated \Mstar~and \SFR~values. Since the subset with \ZSpec~values is dominated by the \QSO~class, it is likely biased toward more massive and luminous host galaxies.}}
\label{fig:SEDProps_GAL_HIST}
\end{figure}

\begin{figure}[t!]
\includegraphics[width=0.48\textwidth]{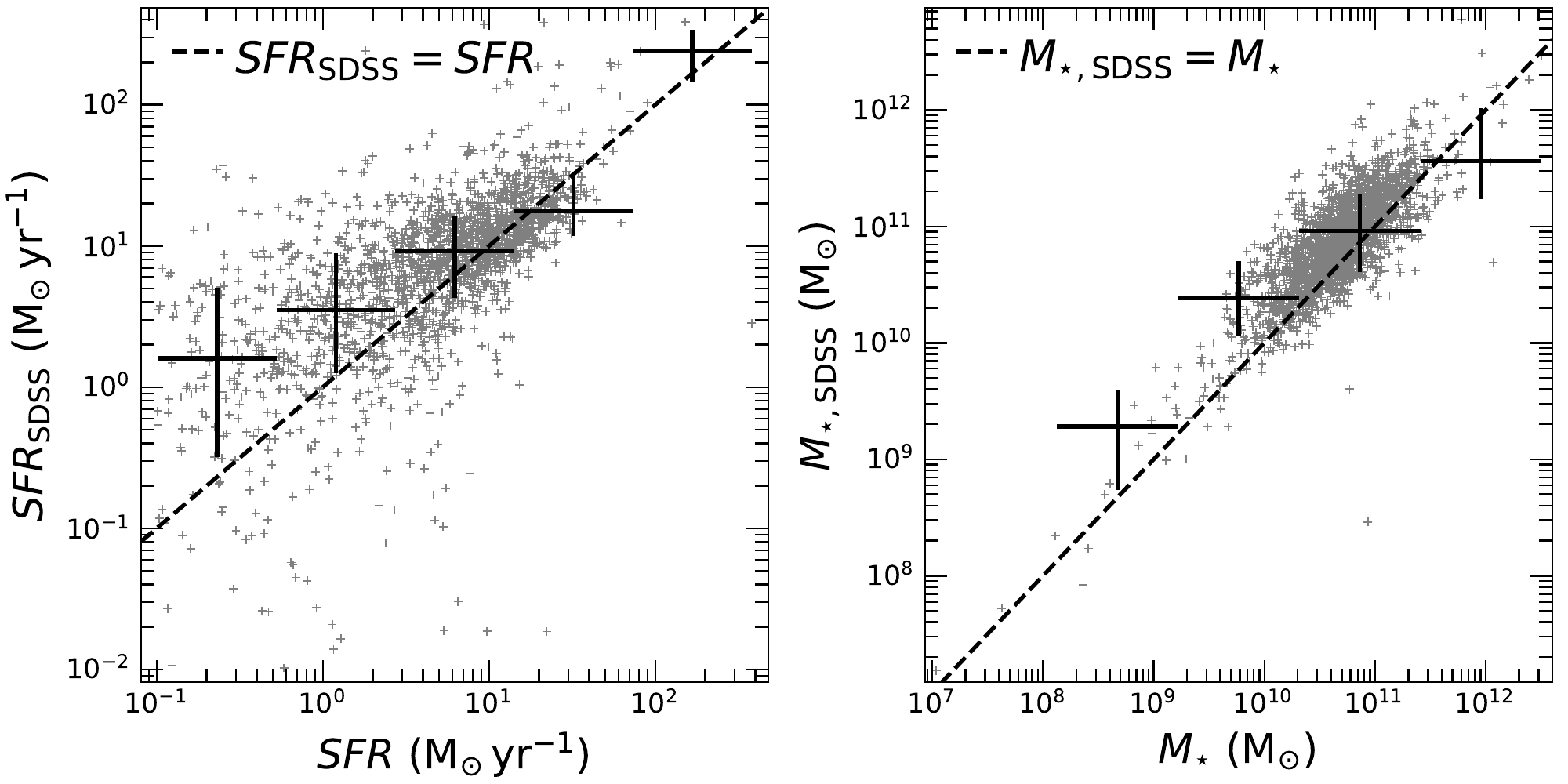} \\
\includegraphics[width=0.48\textwidth]{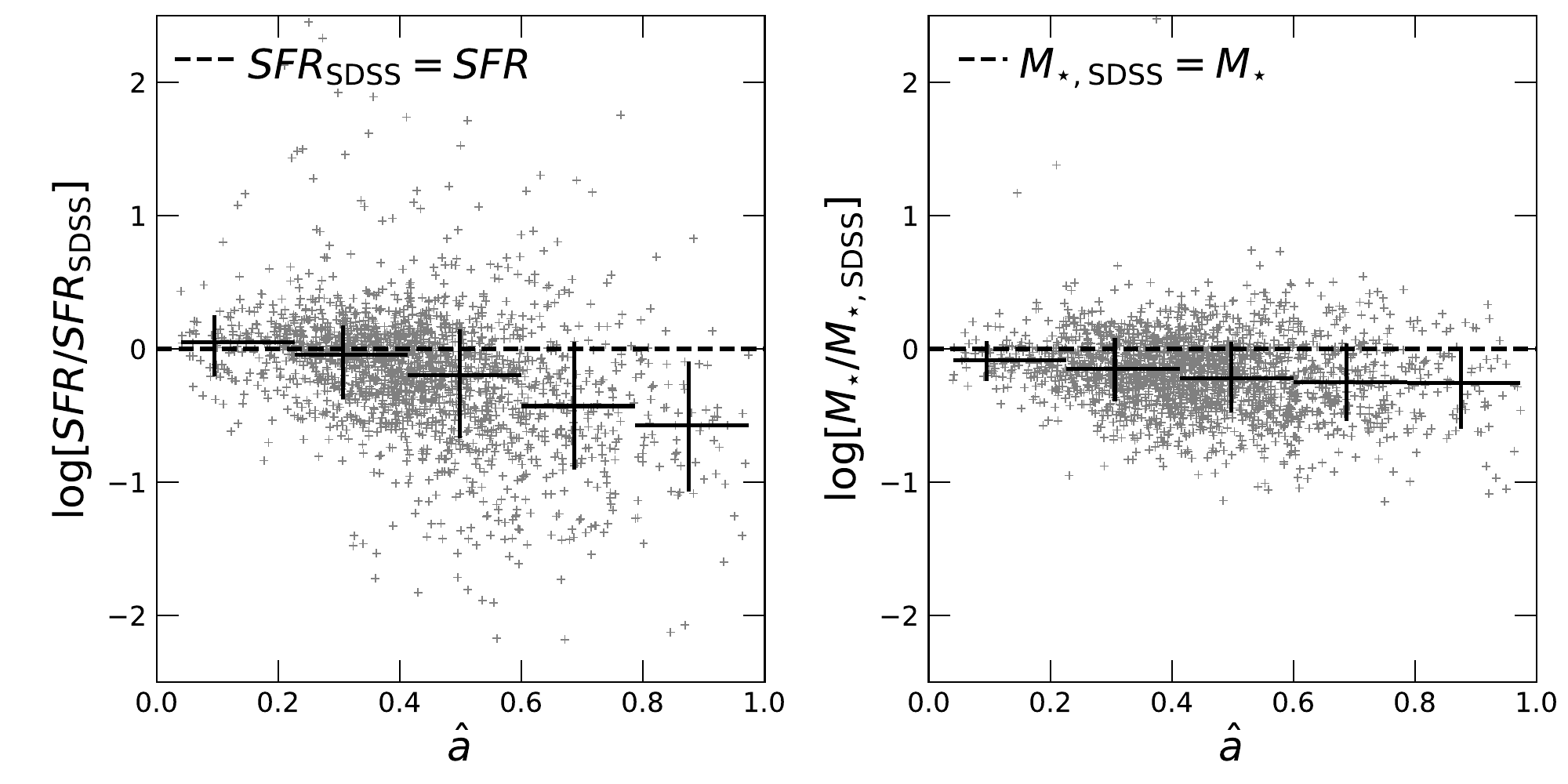}
\caption{\footnotesize{Top: star formation rates (\SFRSDSS; left) and stellar masses (\MstarSDSS; right) from \sdss~DR14 galaxy models plotted against those from our SED modeling procedure (\SFR~and \Mstar). Bottom: ratio between our estimates and those from the \sdss~as a function of the AGN fraction (\ahat). In all panels the black crosses represent median values within logarithmically-spaced bins of even size along the abscissa. Horizontal error bars represent the bin widths while vertical error bars represent the upper and lower 68.3\% bounds. The negative offsets become systematically larger with increasing \ahat~values, consistent with our estimates yielding lower values after removing the AGN component.}}
\label{fig:SFR_MSTAR_CIGALE_SDSS_PLOT}
\end{figure}

The values of \Mstar~and \SFR~are shown in Figure \ref{fig:SEDProps_GAL_HIST}. As with the \LBolAGN~estimates (Section \ref{sec:AGNProps}), the \ZPhot~uncertainties introduce a systematic negative offset in the \Mstar~and \SFR~estimates (median offsets of \MstarSysOffsetPercMedian\%~and \SFRSysOffsetPercMedian\%, respectively). After correcting for these systematic offsets, the subset with only photometric redshifts has smaller \Mstar~values (median of \Mstar\,$=$\,\MstarMedianCorrPhot\,\MSun) and smaller \SFR~values (median of \SFR\,$=$\,\SFRMedianCorrPhot\,\uSFR) compared to the spectroscopic subset (median values of \Mstar\,$=$\,\MstarMedianSpec\,\MSun~and \SFR\,$=$\,\SFRMedianSpec\,\uSFR). This is likely due to the high fraction of QSOs among the spectroscopic subset and hence a bias toward more massive and luminous host galaxies.

\begin{figure}[t!]
\includegraphics[width=0.48\textwidth]{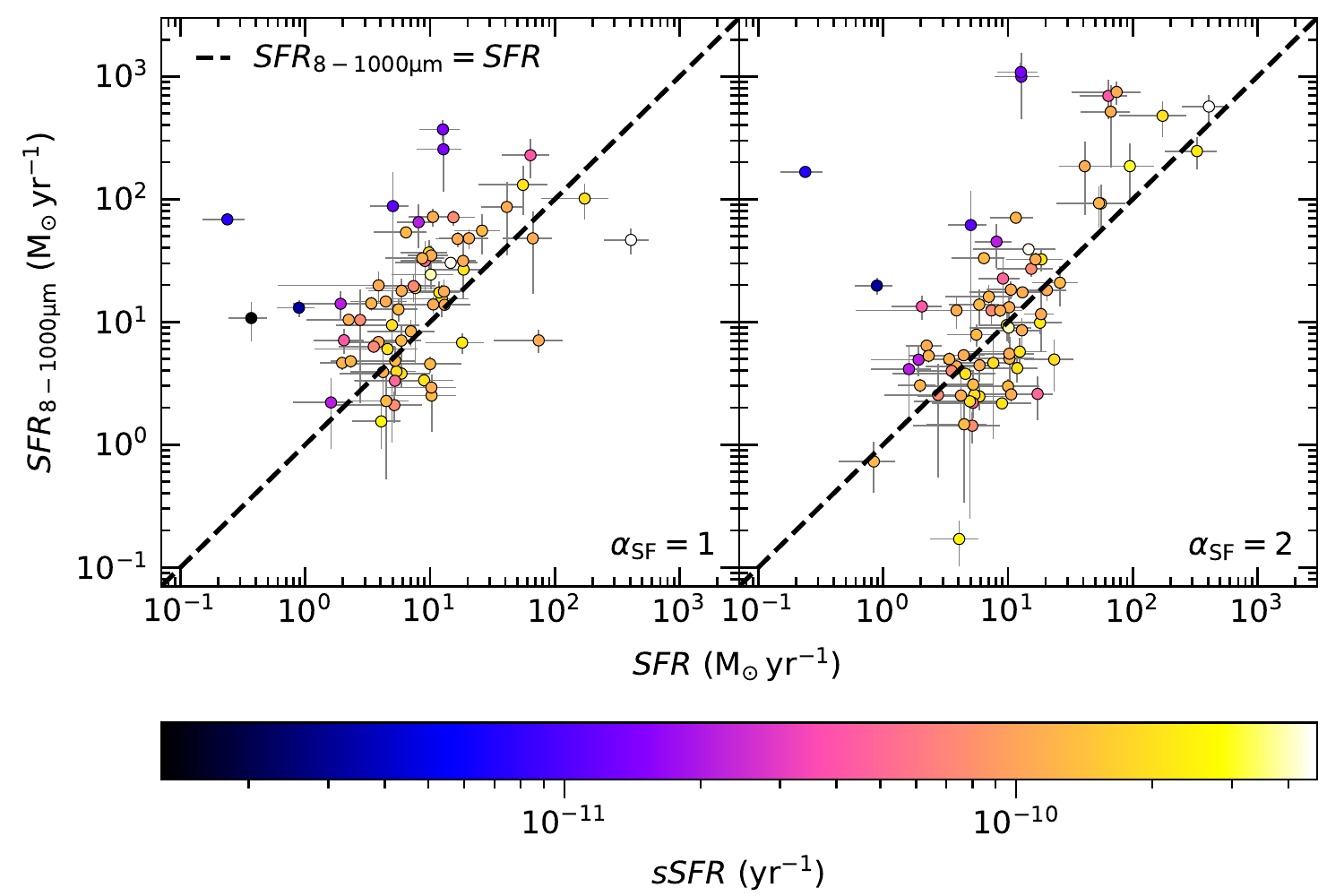}
\caption{\footnotesize{Star formation rates measured from the rest-frame 8$\,-\,$1000\,\micron~luminosity (\SFRFIR) assuming \AlphaSF$\,=1$ (left) and \AlphaSF$\,=2$ (right) plotted against those measured from our SED modeling procedure (\SFR). The error bars represent the upper and lower 68.3\%~bounds. The \SFRFIR~values are over-estimated in systems with low specific \SFRs~(\sSFRs), likely due to the FIR luminosity sensitivity to dust heated by evolved stellar populations.}}
\label{fig:SFR_FIR_PLOT}
\end{figure}

\subsubsection{Comparison with Independent Estimates}
\label{sec:comp}

We compare our \Mstar~and \SFR~estimates with those from the \sdss~DR14 spectroscopic galaxy sample that are based on fitting galaxy models \citep{Conroy:2009} to \sdss~photometry \citep{Montero-Dorta:2016}. The top panels of Figure \ref{fig:SFR_MSTAR_CIGALE_SDSS_PLOT} show that our estimates are systematically offset toward lower values (median offsets of log[\SFR/\SFRSDSS]$\,=\,$\MedianLogSFRRatioCIGALEvSDSS~and log[\Mstar/\MstarSDSS]$\,=\,$\MedianLogMstarRatioCIGALEvSDSS). The offsets are negligible for small AGN fractions (small \ahat~values) but increase in magnitude toward larger \ahat~values (bottom panels of Figure \ref{fig:SFR_MSTAR_CIGALE_SDSS_PLOT}). These negative trends suggest that the systematic offsets are due to our subtraction of an AGN component that is not accounted for in the \sdss~galaxy models. The significantly larger scatter (and the larger offsets) among the low \SFRs~are likely due to these estimates being sensitive to both UV and IR flux that can each have substantial AGN contributions and introduce uncertainties when the star formation signatures are weak.

Since inaccurate AGN subtractions may compromise estimates of host galaxy properties, we have computed independent \SFR~estimates based on FIR (rest-frame 8$\,-\,$1000$\,\micron$) luminosities where AGN contribute significantly less than the host galaxy \citep[e.g.][]{Schartman:2008,Richards2006,Shang:2011,Kirkpatrick:2012,Shi:2013}. We first identify the subset of our sample (we use the spectroscopic subset to avoid biases due to \ZPhot~uncertainties) with photometry from the Photodetector Array Camera and Spectrometer (PACS) on Herschel (bands centered at 70, 100, and 160\,\micron). We then estimate $L_{8\,-\,1000\,\micron}$ by normalizing dust templates \citep{Dale:2002} to the the Herschel/PACS luminosities (minimizing the $\chi_{\mathrm{red}}^2$ statistic to the available detections). The AGN contribution to $L_{8\,-\,1000\,\micron}$ is estimated from the starburst plus AGN composite spectra from \citet{Dale:2014} and the 5$\,-\,$20$\,\micron$ AGN fractions from our SED models.

Figure \ref{fig:SFR_FIR_PLOT} shows the \SFRs~obtained using the dust templates defined by \AlphaSF$\,=\,$1 and 2, where \AlphaSF~parameterizes the distribution of radiation fields contributing to the heated dust mass \citep[typical values for galaxies are \AlphaSF$\,=\,$1\,$-$\,2.5, and only strongly star-forming galaxies have values of \AlphaSF$\,<\,$1; e.g.][]{Dale:2001,Dale:2002}. The star formation rate estimates from the FIR dust templates (\SFRFIR) are correlated with those from our SED modeling, though with a systematic offset toward larger values. The offsets are negligible for systems with higher specific \SFRs~(\sSFRs) and become more significant among those with lower \sSFRs. This result is consistent with the sensitivity of FIR luminosity to dust heated by stars over long timescales that leads to over-estimates for systems with large stellar mass fractions from evolved stars and hence low \sSFRs~\citep[e.g.][]{Kennicutt:2012}.

\subsection{Scattered AGN Light}
\label{sec:scattered}

The AGN template used in our SED models does not directly account for AGN light that has been scattered into the line-of-sight by interactions with electrons and dust \citep[e.g.][]{Antonucci:1985,Kishimoto:1999}. Since the scattered AGN continuum flux rises toward and peaks at UV wavelengths \citep[e.g.][]{Kishimoto:2001,Draine:2003} that trace star formation, the galaxy luminosities returned by our SED models may, in some cases, lead to over-estimated \SFRs. Scattered light is most significant for powerful obscured AGN \citep[e.g.][]{Zakamska:2006}, and indeed some \wise~AGN in the population of hot, dust-obscured galaxies (Hot DOGs) show evidence for excess blue colors due to scattering of AGN light off of dust \citep{Assef:2020}.

Assuming a scattering efficiency at 2800\,\AA~\citep[the monochromatic luminosity at this wavelength is strongly correlated with \SFR; e.g.][]{Madau:1998} of $\epsilon=\,$\UVScatEpsilon~\citep[e.g.][]{Zakamska:2005}, we find that the median value of the scattered-to-stellar flux ratio (defined as the fraction of scattered light relative to the total stellar light for a galaxy) is \ScatToStellarMedian~and generally larger among more luminous AGN. The values extend up to $\sim\,$\ScatToStellarUpper, and therefore UV-based \SFRs~may be over-estimated by up to a factor of several among the most luminous AGN. For reference, we provide the 2800\,\AA~monochromatic AGN luminosities in the catalog.

\begin{deluxetable*}{lll}
\tabletypesize{\footnotesize}
\tablecolumns{3}
\tablecaption{Data Fields of the \WAGC.}
\tablehead{
\colhead{Column Number} &
\colhead{Data Type} &
\colhead{Note}
}
\startdata
1 & ID & Catalog specific unique identifier \\
2 & Name & \wise~source name \\
3 & RA & \wise~source right ascension (degrees) \\
4 & DEC & \wise~source declination (degrees) \\
5 & z\_spec & Spectroscopic redshift (-999 if not available) \\
6 & z\_phot & Photometric redshift \\
7 & z\_best & Best redshift \\
8 & z\_type & Best redshift type [\texttt{spec} or \texttt{phot}] \\
9 & L\_bol\_agn\_best & AGN component bolometric luminosity, best [\uLum] \\
10 & L\_bol\_agn\_16 & AGN component bolometric luminosity, 16th percentile [\uLum] \\
11 & L\_bol\_agn\_84 & AGN component bolometric luminosity, 84th percentile [\uLum] \\
12 & EBV\_best & AGN color excess, best [mag] \\
13 & EBV\_16 & AGN color excess, 16th percentile [mag] \\
14 & EBV\_84 & AGN color excess, 84th percentile [mag] \\
15 & SFR\_best & Host galaxy star formation rate, best [\uSFR] \\
16 & SFR\_16 & Host galaxy star formation rate, 16th percentile [\uSFR] \\
17 & SFR\_84 & Host galaxy star formation rate, 84th percentile [\uSFR] \\
18 & M\_star\_best & Host galaxy stellar mass, best [\MSun] \\
19 & M\_star\_16 & Host galaxy stellar mass, 16th percentile [\MSun] \\
20 & M\_star\_84 & Host galaxy stellar mass, 84th percentile [\MSun] \\
21 & L\_2800\_best & 2800 monochromatic AGN luminosity, best [\uLum] \\
22 & L\_2800\_16 & 2800 monochromatic AGN luminosity, 16th percentile [\uLum] \\
23 & L\_2800\_84 & 2800 monochromatic AGN luminosity, 84th percentile [\uLum] \\
24 & L\_E\_best & Elliptical component bolometric luminosity, best [\uLum] \\
25 & L\_E\_16 & Elliptical component bolometric luminosity, 16th percentile [\uLum] \\
26 & L\_E\_84 & Elliptical component bolometric luminosity, 84th percentile [\uLum] \\
27 & L\_Sbc\_best & Spiral component bolometric luminosity, best [\uLum] \\
28 & L\_Sbc\_16 & Spiral component bolometric luminosity, 16th percentile [\uLum] \\
29 & L\_Sbc\_84 & Spiral component bolometric luminosity, 84th percentile [\uLum] \\
30 & L\_Im\_best & Irregular component bolometric luminosity, best [\uLum] \\
31 & L\_Im\_16 & Irregular component bolometric luminosity, 16th percentile [\uLum] \\
32 & L\_Im\_84 & Irregular component bolometric luminosity, 84th percentile [\uLum] \\
33-49 & f\_i\_mod & Model fluxes, $i=1-17$ [Jy] \\
50-66 & k\_i\_mod & Photometric k-corrections, $i=1-17$ \\
67-83 & f\_i\_obs & Observed fluxes (-999 if no detection), $i=1-17$ [Jy] \\
84-100 & f\_e\_i\_obs & Observed flux errors (-999 if no detection), $i=1-17$ [Jy]
\enddata
\tablecomments{Columns $1-4$ are basic source descriptors, columns $5-8$ are described in Section \ref{sec:redshifts}, columns $9-14$ are described in Section \ref{sec:AGNProps}, columns $15-20$ are described in Section \ref{sec:GalProps}, columns $21-23$ are described in Section \ref{sec:scattered}, columns $24-66$ are additional output values related to the SED fitting process, and columns $67-100$ are the input photometric fluxes and their errors.}
\label{tab:cat}
\end{deluxetable*}

\subsection{Catalog Contents and Structure}
\label{sec:Cat}

The primary data products in the \WAGC~we present here are photometric redshifts (and spectroscopic redshifts, if available), AGN physical properties (bolometric luminosities and nuclear obscuration), and host galaxy physical properties (stellar masses and \SFRs). The catalog also includes bolometric luminosities of the individual galaxy components (elliptical, Sbc, and irregular) and the 2800\,\AA~monochromatic AGN luminosities. The distance properties are computed assuming the cosmology stated in Section \ref{sec:intro} and the best available redshifts. For completeness, the catalog includes observed and model photometric fluxes for each filter from the full photometric data set plus the model $k-$corrections. Table \ref{tab:cat} lists the full suite of data included in the catalog. Uncertainties on all derived parameters are estimated by re-fitting each SED 1,000 times with random Gaussian noise (standard deviations equal to the photometry 1$\sigma$ uncertainties) added. Uncertainties in the host galaxy model fluxes are propagated into the stellar mass and \SFR~uncertainties.

\begin{figure}[t!]
\includegraphics[width=0.48\textwidth]{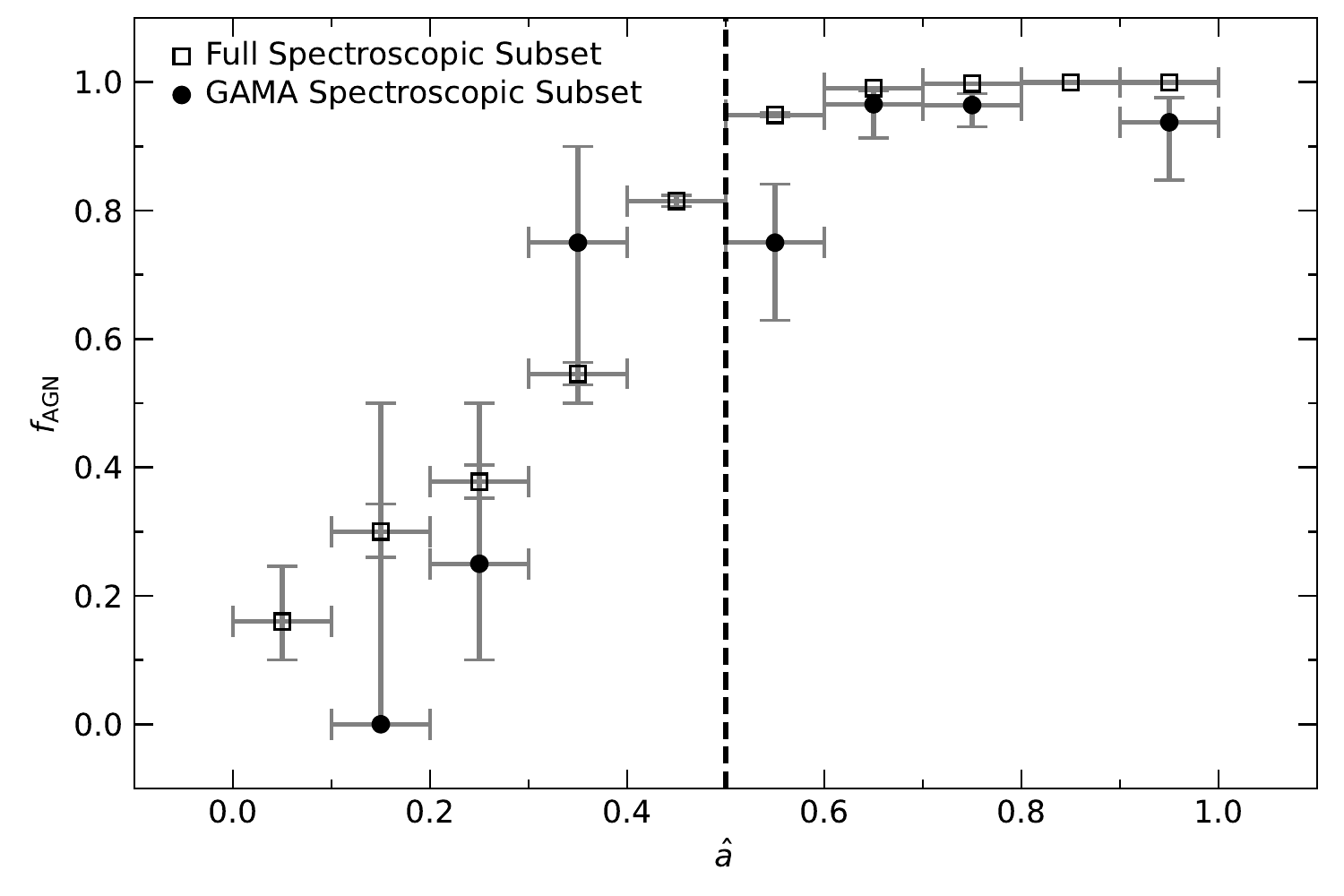}
\caption{\footnotesize{Fraction of \wise~AGN candidates with spectroscopic AGN signatures (\AGNFrac) for the full spectroscopic subset (open squares) and the subset within the \gama~survey (filled circles) as a function of the AGN fraction (from our SED models; \ahat). The vertical error bars denote the upper and lower 68.3\%~binomial uncertainties, and the horizontal error bars denote the bin width. The threshold of \ahat$\,>\,$\AHatThresh~yields spectroscopic AGN fractions of \HiAHatAGNPerc$\%$ and \GAMAHiAHatAGNPerc$\%$ for the full and \gama~spectroscopic subsets, respectively.}}
\label{fig:AGN_FRAC_AHAT_PLOT}
\end{figure}

\section{Contamination from Purely Star-Forming Galaxies}
\label{sec:contamination}

\begin{figure}[t!]
\includegraphics[width=0.48\textwidth]{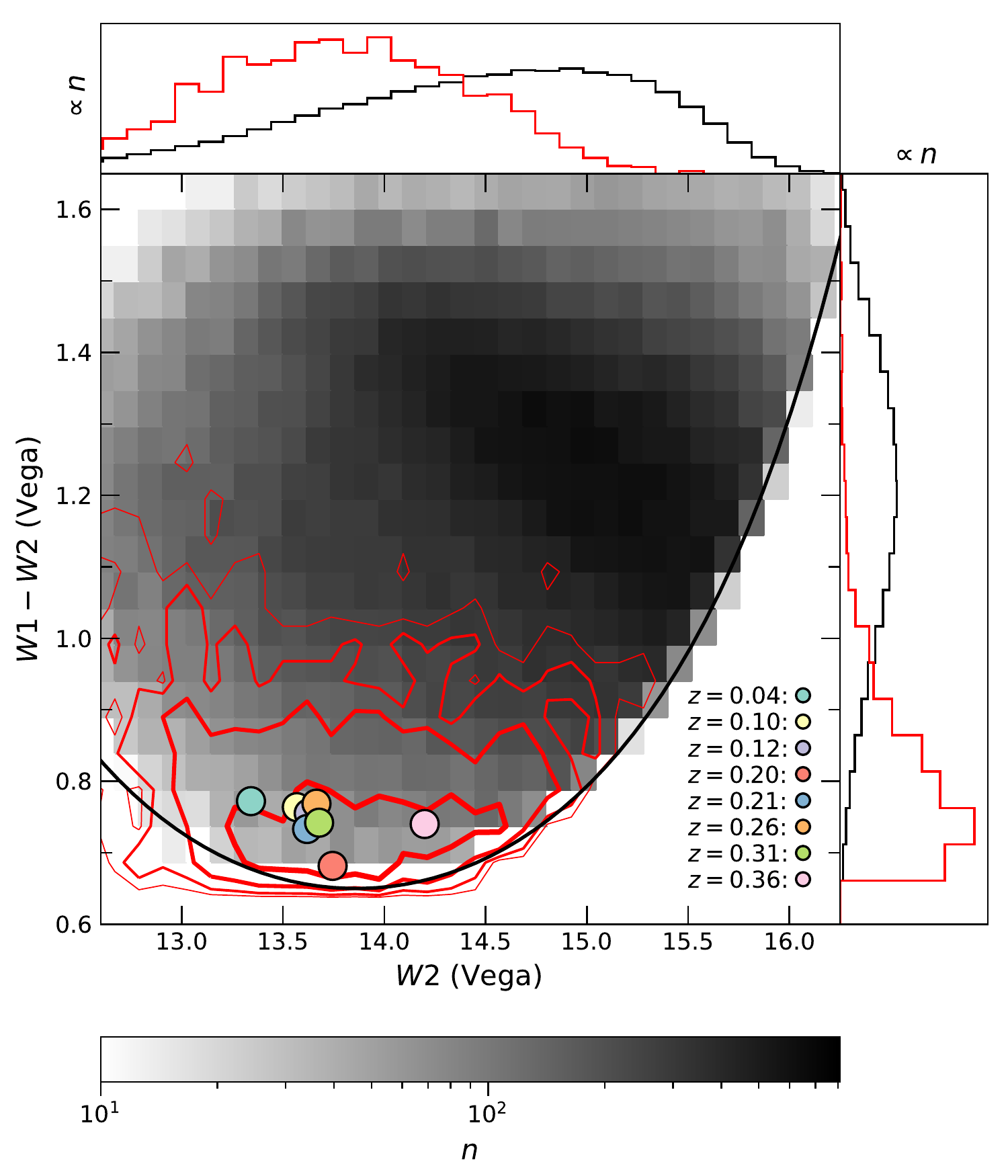}
\caption{\footnotesize{\wiseonetwo~versus \wisetwo~for \wise~AGN candidates in our catalog. The \AssefCutCol~line indicates the function used to select AGN candidates for the \R~version of the \WAC. Number densities are indicated by the \HiAHatScaleCol-scale pixels for the AGN and by the \LoAHatContourCol~contours for the contaminants. Histograms for the X- and Y-axis values are shown along the top and left, respectively, for the AGN (\HiAHatHistCol) and contaminants (\LoAHatContourCol), each normalized to an integrated value of unity. Contaminants are preferentially seen at bluer \wiseone$\,-\,$\wisetwo~colors and brighter \wisetwo~magnitudes relative to AGN. Individual contaminants with the lowest $\chi_{\mathrm{red}}^2$ pure galaxy SED models in eight~evenly-spaced redshift bins over the range \z$\,=\,$\ZBinsCigaleLoLim$\,-\,$\ZBinsCigaleHiLim~are shown as colored circles. The model fits for these individual galaxies are shown in Figure \ref{fig:CIGALE_MODELS}.}}
\label{fig:Assef2018_AHAT_PLOT}
\end{figure}

In this section we identify and discuss the nature of contaminants in our catalog. We use the \sdss~spectroscopic subset to identify sources from our catalog with AGN signatures based on either 1) a \QSO~classification or 2) a \GAL~classification plus optical emission line ratios consistent with Seyfert galaxies based on all three diagnostics in \citet{Kewley:2006}. In Section \ref{sec:cont_frac} we quantify the contamination fraction, in Section \ref{sec:cont_colors} we identify the regions of redshift and \wise~color-space where the contaminants are found, and in Section \ref{sec:cont_nature} we discuss their physical nature.

\begin{figure*}[t!]
\hspace*{0.001in} \includegraphics[width=0.99\textwidth]{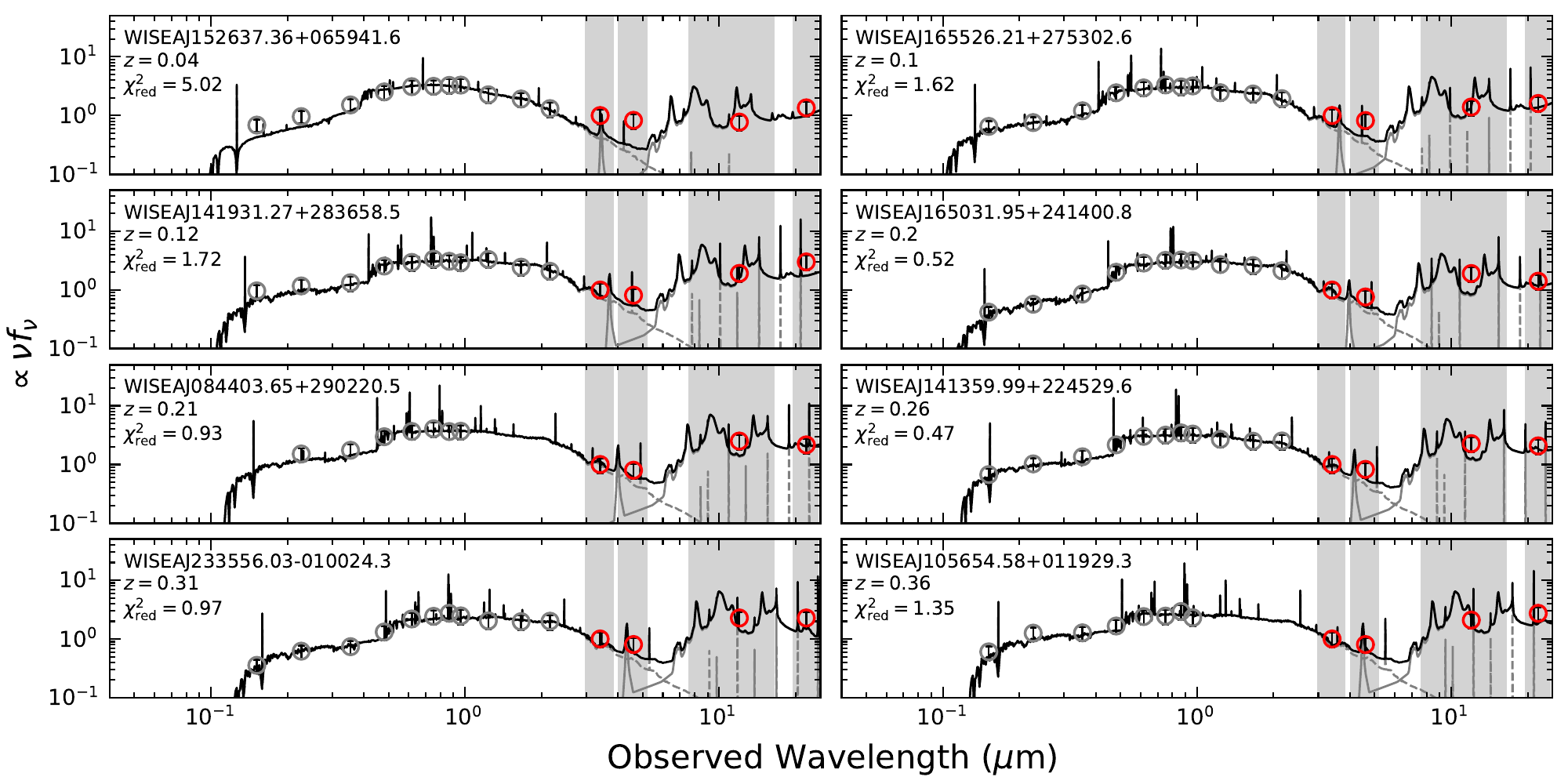}
\caption{\footnotesize{SEDs and galaxy models (normalized at 3.6\,\micron) for contaminants with the lowest reduced $\chi^2$ ($\chi_{\mathrm{red}}^2$) statistic in eight evenly-spaced redshift bins over the range \z$\,=\,$\ZBinsCigaleLoLim$\,-\,$\ZBinsCigaleHiLim. The \wise~filter FWHMs are shown as gray-shaded regions, and the \wise~photometry is shown in red. The attenuated stellar-plus-nebular emission and dust emission are indicated by the dashed and solid gray lines, respectively, while the best-fit model sum is shown as a solid black line. The best model fits are obtained for galaxies at redshifts $z=0.2-0.3$.}}
\label{fig:CIGALE_MODELS}
\end{figure*}

\subsection{The Contamination Fraction}
\label{sec:cont_frac}

The fraction of sources with spectroscopic AGN signatures is \ReliableFrac\% and is positively correlated with \ahat~(Figure \ref{fig:AGN_FRAC_AHAT_PLOT}). Those with \ahat$\,>\,$\AHatThresh~have an average AGN fraction of \HiAHatAGNPerc\%. Since the spectroscopic subset is subject to \sdss~selection effects, in Figure \ref{fig:AGN_FRAC_AHAT_PLOT} we also show the subset within the \gamatitle~(\gama) survey that is complete to a magnitude limit of $r=19.8$ \citep{Baldry:2010}. The AGN fraction in the \gama~subset is \GAMAReliableFrac\%~and has a positive dependence on \ahat~that is consistent with the full spectroscopic sample when accounting for the binomial uncertainties. The \gama~subset with \ahat$\,>\,$\AHatThresh~has an average AGN fraction of \GAMAHiAHatAGNPerc\%. 

We remark that these spectroscopic AGN identifications are biased toward unobscured AGN and therefore likely provide an underestimate of the AGN fraction. While narrow emission line diagnostics may in some cases yield inaccurate AGN classifications \citep[e.g.][]{Wylezalek:2018}, this effect will be negligible since the vast majority of spectroscopic AGN signatures in our catalog are based on the \QSO~classification.

\subsection{The \wise~Colors and Redshifts of Contaminants}
\label{sec:cont_colors}

The locations of spectroscopic AGN and contaminants in the color-magnitude space of the \WAC~are shown in Figure \ref{fig:Assef2018_AHAT_PLOT}. Contaminants are biased toward bluer \wiseone$\,-\,$\wisetwo~colors (median of \wiseone$\,-\,$\wisetwo$\,=\,$\WOneTwoMedianLoAHat) relative to AGN (median of \wiseone$\,-\,$\wisetwo$\,=\,$\WOneTwoMedianHiAHat), likely due to MIR spectral slopes being steeper among AGN than among purely star-forming galaxies \citep[e.g.][]{Donley:2012,Chang:2017}. The fainter magnitudes among the AGN (median of \wisetwo$\,=\,$\WTwoMedianHiAHat) compared to contaminants (median of \wisetwo$\,=\,$\WTwoMedianLoAHat) may be due, at least in part, to the bias of \sdss~QSOs toward higher redshifts (e.g. Figure \ref{fig:Z_HIST}).

To understand how purely star-forming galaxies can pass the \wise~AGN selection criteria as a function of redshift, we have fit the SED of each contaminant with a pure galaxy model using \cigale~and the same configuration described in Section \ref{sec:SFR}. To identify the contaminants that are best-fit within the constraints of our models, in each of eight evenly-spaced redshift bins over the range \z$\,=\,$\ZBinsCigaleLoLim$\,-\,$\ZBinsCigaleHiLim~(corresponding to the redshift range of spectroscopic contaminants in our sample) we select the fit with the lowest $\chi_{\mathrm{red}}^2$ value. Their locations within the \wise~AGN color-magnitude space are marked in Figure \ref{fig:Assef2018_AHAT_PLOT} and indicate that they are representative of the overall contaminant population. Their SEDs and best-fit models (Figure \ref{fig:CIGALE_MODELS}) show how the \wiseonetwo~colors are affected by a combination of (attenuated) stellar continuum and dust that includes several strong PAH emission lines that migrate through the \wise~filter set with increasing redshift. The best fits are achieved at redshifts of \z\,$=$\,0.2\,$-$\,0.3 when the \wisethree~filter aligns with a local dip in the redshifted dust continuum and allows for larger dust contribution to the \wisetwo~filter.

The redshift distributions of the AGN and contaminants are shown in the top panel of Figure \ref{fig:Z_AHAT_SSFR_MSTAR_OFFSET_PLOT}. The redshifts of contaminants have a median value of \z$\,=\,$\LoAHatAllZMedian~that is consistent with the redshifts corresponding to the best pure galaxy SED models (Figure \ref{fig:CIGALE_MODELS}), suggesting that the majority of contamination in the \WAC~is due to dust heated by star formation in galaxies at \z\,$=$\,0.2\,$-$\,0.3 and with relatively blue \wiseonetwo~colors. While high-redshift (\z$\,\sim\,$$1-2$) galaxies are another potential source of contamination (Section \ref{sec:intro}) they are unlikely due to the \wisetwo~magnitude dependence.

\subsection{The Nature of the Contaminants}
\label{sec:cont_nature}
 
\begin{figure}[t!]
\includegraphics[width=0.48\textwidth]{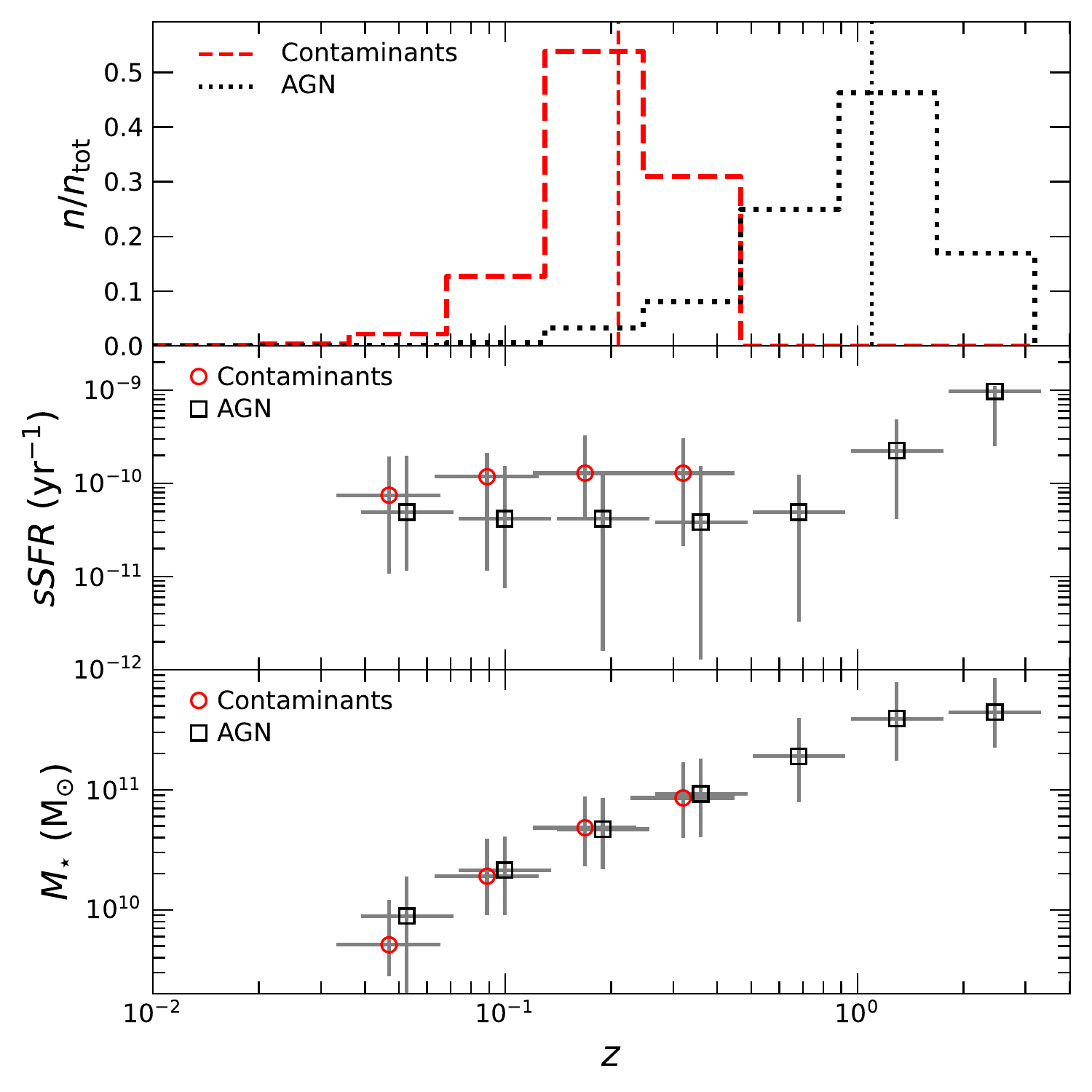}
\caption{\footnotesize{Top: distribution of redshifts (\z) for contaminants (red, dashed) and AGN (black, dotted), each normalized to an integrated value of unity. The vertical lines of corresponding color and style indicate the median values. The median value of the contaminants (\z$\,=\,$\LoAHatAllZMedian) is consistent with the redshift range of the best pure galaxy models ($z=0.2-0.3$; Figure \ref{fig:CIGALE_MODELS}). Middle and bottom: specific \SFRs~(\sSFR) and stellar masses (\Mstar), respectively, as a function of \z~for the contaminants (red circles) and AGN (black squares). The data points represent median values within logarithmically-spaced bins of even size along the abscissa, horizontally offset for clarity. Horizontal error bars represent the bin widths while vertical error bars represent the upper and lower $68.3\%$ bounds. The \sSFRs~of the contaminants are systematically elevated relative those of the AGN, while no systematic offset of the \Mstar~values is observed.}}
\label{fig:Z_AHAT_SSFR_MSTAR_OFFSET_PLOT}
\end{figure}

Purely star-forming galaxies are theorized to pass \wise~color qualifications of AGN selection due to large ionization parameters from the stellar radiation field and significant reservoirs of heated dust \citep{Hainline:2016, Satyapal:2018}, and these conditions are most likely to exist in galaxies with large \sSFRs. In the middle panel of Figure \ref{fig:Z_AHAT_SSFR_MSTAR_OFFSET_PLOT} we plot \sSFRs~of the AGN and contaminants as a function of redshift. In each redshift bin occupied by contaminants, the \sSFRs~of contaminants are consistent with those of AGN within their respective uncertainties. However, the \sSFRs~of contaminants are systematically elevated, relative to those of AGN, and suggest that enhanced \SFRs~(per unit stellar mass) contribute to the red \wiseone$\,-\,$\wisetwo~colors observed in galaxies without spectroscopic AGN signatures. 

Contamination from low-mass galaxies (\Mstar\,$<10^{9.5}$\,\MSun) without AGN may also be significant \citep{Hainline:2016, Satyapal:2018,Lupi:2020} since they are frequently star-forming and are known to have low metallicities that result in large effective stellar temperatures and harder radiation fields for heating dust \citep[e.g.][]{Griffith:2011,Remy-Ruyer:2015,O'Connor:2016}. However, while lower metallicities result in hotter dust, they also correspond to lower overall dust abundances that make achieving the red \wiseone$\,-\,$\wisetwo~colors difficult \citep{Satyapal:2018}. Indeed, the bottom panel of Figure \ref{fig:Z_AHAT_SSFR_MSTAR_OFFSET_PLOT} shows no evidence for systematically lower stellar masses among the contaminants.

\begin{figure*}[t!]
\includegraphics[width=\textwidth]{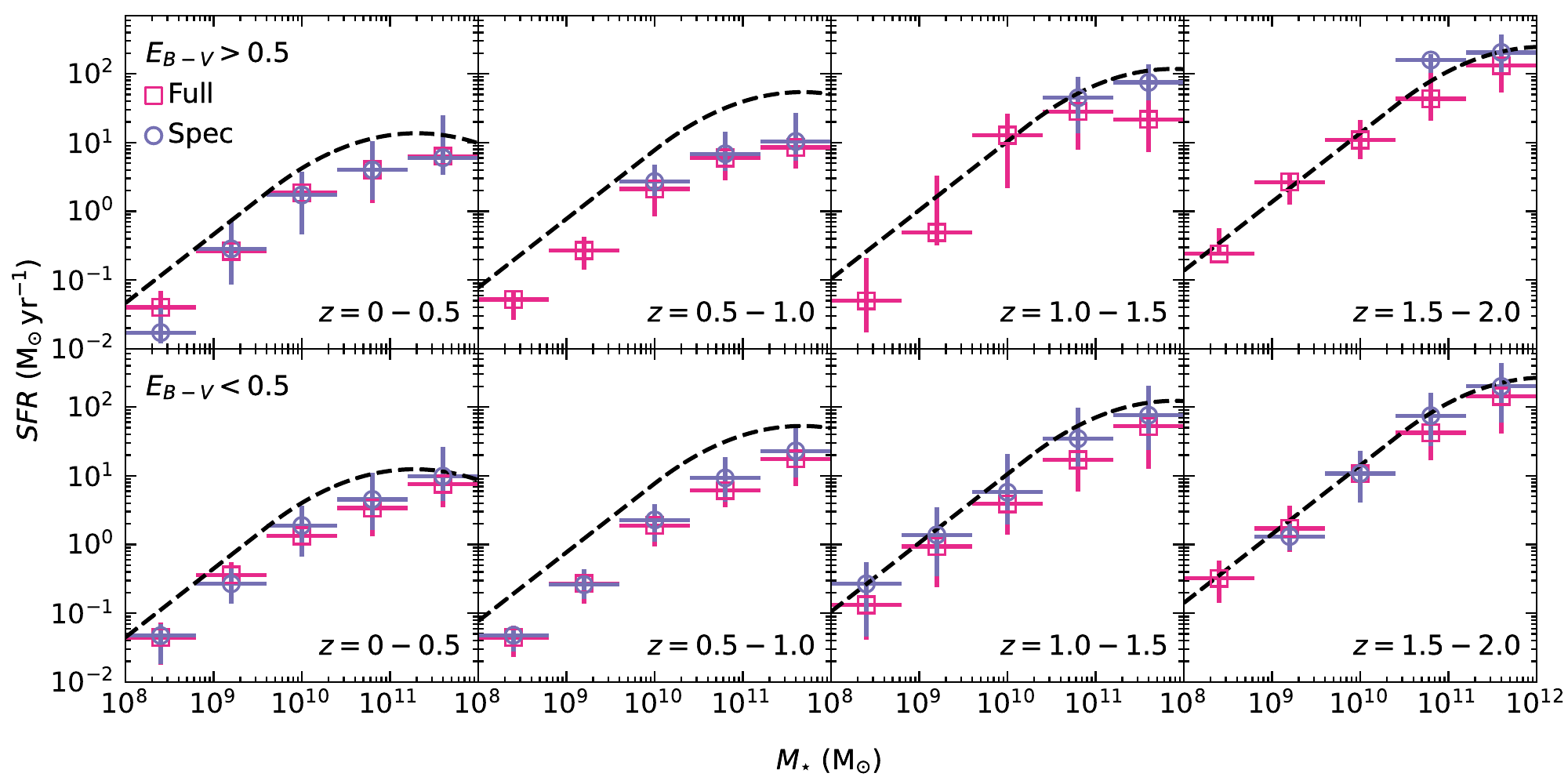}
\caption{\footnotesize{Host galaxy star formation rate (\SFR) as a function of host galaxy stellar mass (\Mstar) for obscured (\EBV$\,\ge0.5$; top) and unobscured (\EBV$\,<0.5$; bottom) \wise~AGN (restricted to sources with \ahat$\,>\,$\AHatThresh). In each redshift bin the dashed line shows the redshift-dependent relation between \SFR~and \Mstar~from \citet{Schreiber:2015} where we have used the median redshift value in that bin. The magenta squares and purple circles represent median values within logarithmically-spaced bins of even size along the abscissa for the full sample and the subset with spectroscopic redshifts, respectively. Horizontal error bars represent the bin widths while vertical error bars represent the upper and lower $68.3\%$ bounds. Toward lower redshifts the \wise~AGN host galaxies show increasing systematic offsets toward lower specific \SFRs~compared to the relation for star-forming galaxies.}}
\label{fig:MSTAR_SFRUV_PLOT}
\end{figure*}

\section{Co-evolution of \wise~AGN and Their Host Galaxies}
\label{sec:Evol}

To investigate the evolutionary histories of \wise~AGN host galaxies, we use our catalog to examine the relationship between \SFR~and \Mstar~separately for obscured AGN (\EBV$\,\ge0.5$; galaxy-dominated SED continua) and unobscured AGN (\EBV$\,<0.5$; AGN-dominated SED continua) in Figure \ref{fig:MSTAR_SFRUV_PLOT}. To avoid contribution from potential contaminants (Section \ref{sec:contamination}), we omit galaxies with \ahat$\,<\,$\AHatThresh. To test the impact of photometric redshift uncertainties, we also show the subset with spectroscopic redshifts. Since the normalization of the function relating \SFR~and \Mstar~increases with redshift for star forming galaxies \citep[e.g.][]{Daddi:2007,Rodighiero:2010,Karim:2011,Whitaker:2012,Schreiber:2015,Spilker:2018}, we plot these values in four evenly-spaced redshift bins over the range \z$\,=\,$$0-2$ (where the sample is well-represented). 

Our sample of AGN host galaxies is relatively shallow compared to those built from deeper surveys such as COSMOS \citep[e.g.][]{Suh:2017,Suh:2019} and therefore biased toward more massive galaxies at each epoch. Regardless, similar to previous studies of AGN hosts \citep[e.g.][]{Boyle:1998,Chapman:2005,Santini:2012,Azadi:2015}, we observe a trend of increasing \SFR~and \Mstar~values with redshift that reflects the increasing cold gas fractions \citep[e.g.][]{Boselli:2001}. 

The largest \SFRs~are observed in massive galaxies at high redshifts and have values similar to those seen for powerful X-ray AGN \citep[e.g.][]{Rodighiero:2010,Rosario:2012,Bernhard:2016}. These are likely gas-rich galaxies hosting luminous AGN and that may have experienced mergers similar to those observed for ULIRGs \citep[e.g.][]{Sanders:1988}. Indeed, the systems with the highest \SFRs~also have relatively elevated values of \EBV~suggesting they may be dust-rich systems.

Toward lower redshifts, the \sSFRs~show increasing systematic negative offsets from the prediction for star-forming galaxies, reaching offsets of \MeanSFRAllOffsetObscPercA\%~and \MeanSFRAllOffsetUnObscPercA\%~for the obscured and unobscured samples, respectively. As seen in Figure \ref{fig:MSTAR_SFRUV_PLOT}, these offsets are qualitatively the same (accounting for the uncertainties) when limited to the sample with spectroscopic redshifts. These results may suggest that, once the gas has reached the SMBH and triggered an AGN, most of the available cold gas has been used to form stars so that the galaxies move toward a passively evolving sequence at low redshifts \citep[e.g.][]{Peng:Y:2010,Renzin:2015,Spilker:2018,McPartland:2019}. Such quiescent galaxies are known to be the dominant hosts of QSOs at lower redshifts based on imaging \citep[e.g.][]{Taylor:1996,McLure:1999} and spectroscopy \citep[e.g.][]{Nolan:2001}. Moreover, this trend may be similar to that seen for hard X-ray selected AGN at $z<0.05$ \citep[e.g.][]{Shimizu:2015}, for optically-selected AGN \citep[e.g.][]{Leslie:2016}, and possibly for the hosts of many well-studied low-redshift Seyfert galaxies with relatively red colors \citep[e.g.][]{Kotilainen:1994}.
 
These results are overall consistent with evolutionary models in which the bulk of star formation occurs before the peak in SMBH accretion and associated onset of a QSO phase \citep[see][and references therein]{Hickox:2009}. Interestingly, this result also suggests that, while \wise~AGN color selection is less sensitive to obscuration than optical or even X-ray selection, the requirement that the intrinsic AGN component represent a significant fraction of the source emission (i.e. \ahat$\,>\,$\AHatThresh) introduces a bias toward luminous and powerful AGN that are triggered after the global cold gas supply for star formation has been depleted. This hypothesis suggests a delay between star formation and AGN triggering that is observed in galaxies with low gas fractions \citep[e.g.][]{Kaviraj:2015b,Kaviraj:2015a,Shabala:2017}. Alternatively, if significant star formation is still occurring when the AGN is triggered, then negative AGN feedback may also contribute to lower \sSFRs~in nearby galaxies \citep[e.g.][]{Schawinski:2007}.

\section{The Connection Between Nuclear Obscuration and SMBH Growth}
\label{sec:SMBHGrowth} 

While large reservoirs of circum-nuclear dust and gas provide material for accretion onto SMBHs, negative feedback can remove obscuring material \citep[as is assumed in AGN evolutionary models; e.g.][]{Hopkins:2009,Hickox:2009}. Several studies have presented a dynamic picture in which the obscured AGN fraction increases with AGN luminosity among low luminosity AGN \citep[e.g.][]{Beckmann:2009,Burlon:2011,Kawamuro:2016} but then declines toward more luminous AGN \citep[the `receding torus' model; e.g.][]{Lawrence:1991,Mulchaey:1992,Granato:1994,Ueda:2003,Simpson:2005,Lusso:2013,Assef:2013,Merloni:2014}.

Since AGN obscuration traces the amount of dust near the central SMBH, we use the SED-based measurements of \EBV~in our catalog to investigate its direct physical connection with both AGN bolometric luminosity and Eddington ratio. The Eddington ratio is defined as \FEdd$\,=\,$\LBolAGN$\,/\,$\LEdd, and Eddington luminosities are computed as \LEdd$\,=4\pi c\,G$\MBH$\mu_{e}/\sigma_{T}$, where $G$ is the gravitational constant, $\mu_{e}$ is the mass per unit electron, and $\sigma_{T}$ is the Thomson scattering cross-section \citep{Krolik:1999}. We only do this analysis for the spectroscopic QSO subset with SMBH mass (\MBH) estimates \citep{Rakshit:2020} based on the 5100\,\AA~monochromatic luminosity, the H$\beta$ broad emission line FWHM, and the mass-scaling relation from \citet{Assef:2011}. Since the AGN template used in our modeling is relatively blue \citep[corresponding to the composite QSO template from \citealt{Richards2006} with \EBV$\,\sim$\,0.05;][]{Assef2010}, for this analysis we subtract \EBV$\,=0.05$ from our obscuration measurements.

\begin{figure}[t!]
\includegraphics[width=0.48\textwidth]{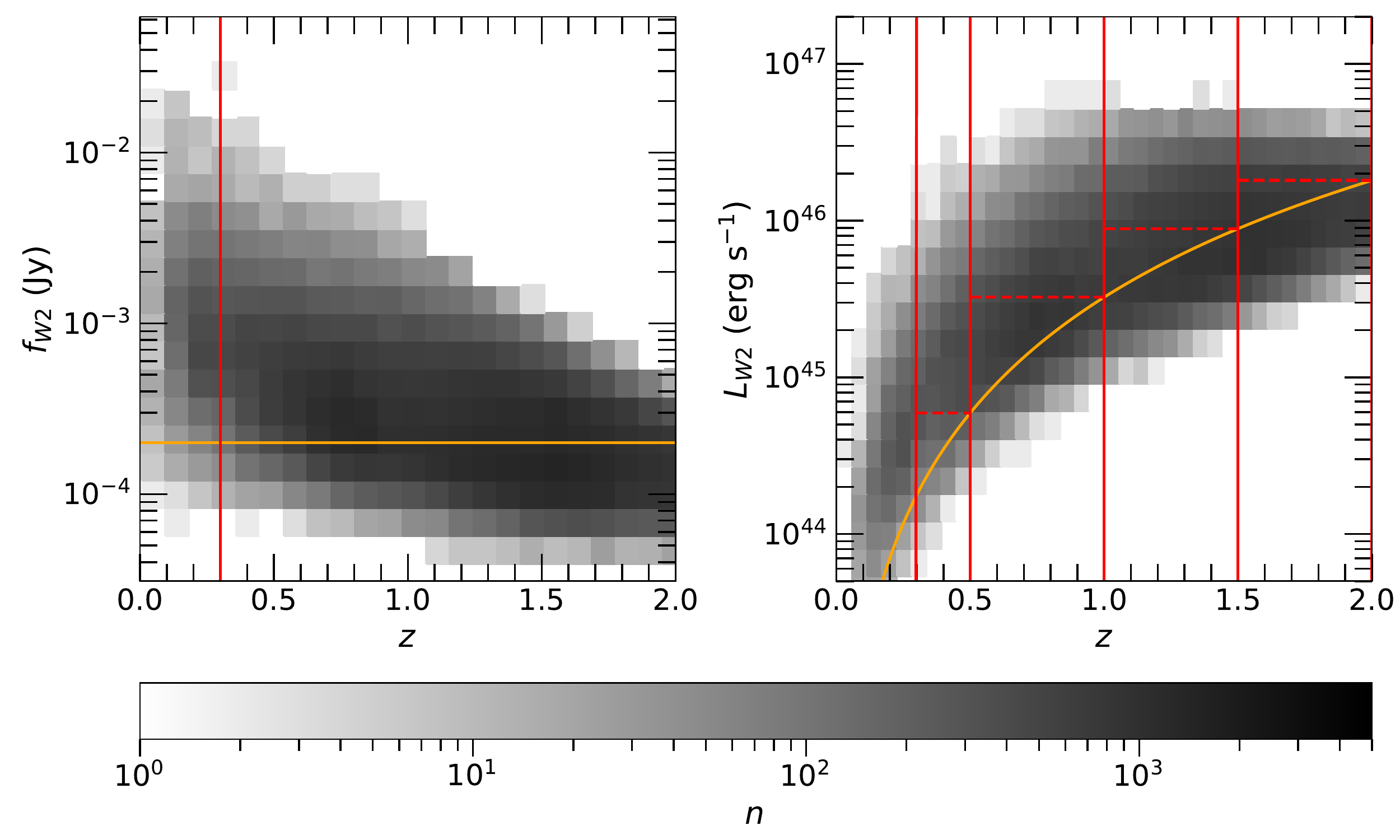}
\caption{\footnotesize{\wisetwo~flux ($f_{W2}$; left) and luminosity ($L_{W2}$; right) as a function of host galaxy redshift (\z) for \wise~AGN in our catalog with the spectroscopic classification of \QSO. Left: the horizontal orange line indicates our adopted flux lower limit ($f_{W2}>\,$\FWTwoLoLim$\,$Jy) that is chosen to approximately align with the faintest sources at a redshift of $z=0.3$ (vertical red line). Right: the curved orange line represents luminosities corresponding to the flux lower limit as a function of redshift. The vertical red lines correspond to the redshift bin edges, and the horizontal red, dashed lines indicate the \wisetwo~luminosity lower limits in each bin. Sources above these lower luminosity limits are used in Figure \ref{fig:EBV_LBOLAGN_FEDD_PLOT}.}}
\label{fig:LBOLAGN_Z_PLOT}
\end{figure}

\begin{figure}[t!]
\includegraphics[width=0.48\textwidth]{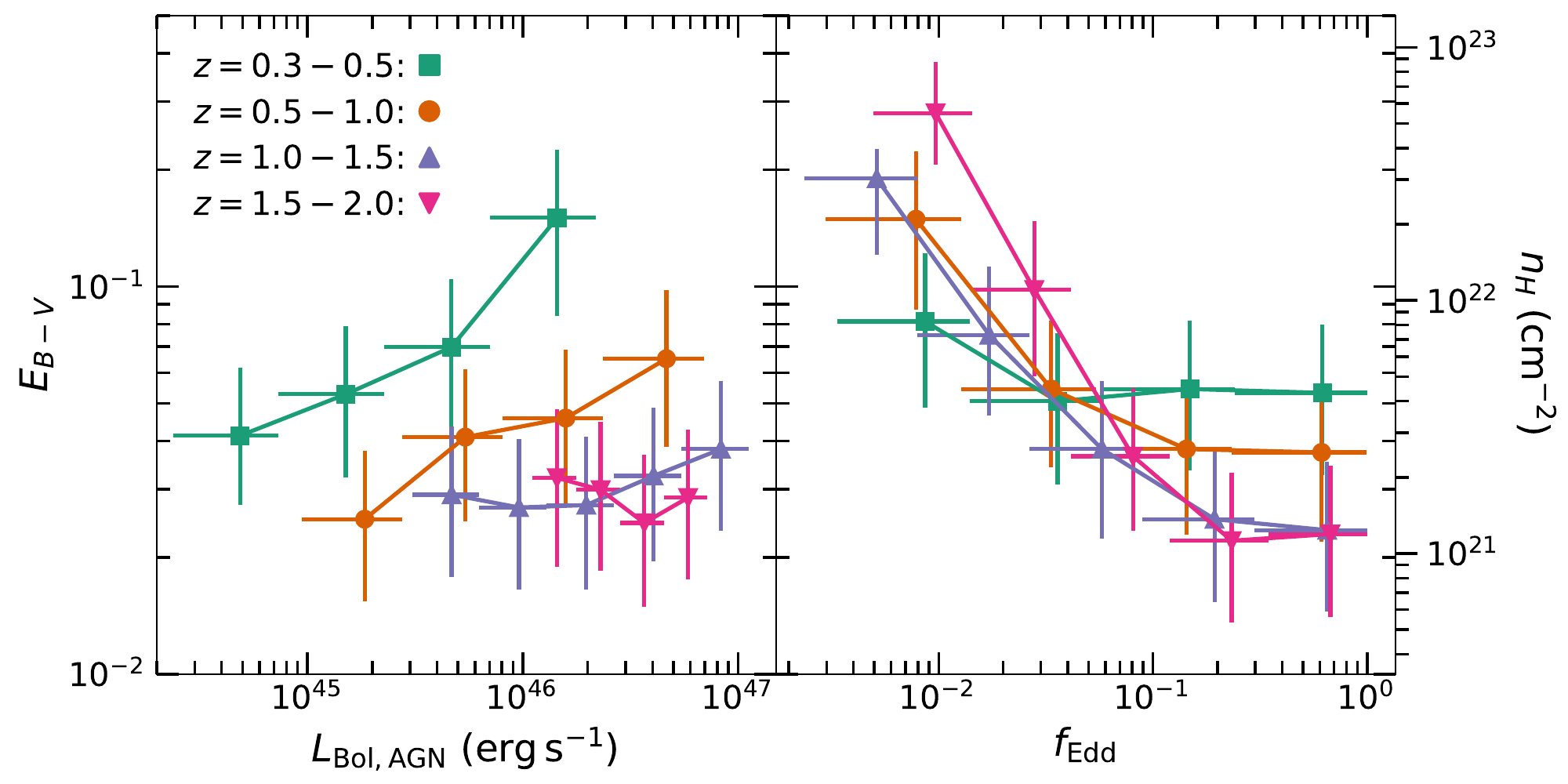}
\caption{\footnotesize{Color excess (\EBV) as a function of AGN bolometric luminosity (\LBolAGN; left) and Eddington ratio (\FEdd; right) for all sources in the redshift bins defined in Figure \ref{fig:LBOLAGN_Z_PLOT}. The data points in each redshift bin represent median values within logarithmically-spaced bins of even size along the abscissa. Horizontal error bars represent the bin widths while vertical error bars represent the upper and lower $68.3\%$ bounds. The equivalent values of extra-galactic column density \citep[\NH; using the gas-to-dust conversion from][]{Maiolino:2001} are shown on the far right axis. \EBV~shows a slight positive correlation with \LBolAGN~that declines in significance toward higher redshifts. On the other hand, \EBV~is negatively correlated with \FEdd, with the significance increasing toward higher redshifts. These results are consistent with the `receding torus' model and suggest that the obscuring dust is within the SMBH sphere of gravitational influence.}}
\label{fig:EBV_LBOLAGN_FEDD_PLOT}
\end{figure}

To account for the strong dependence of AGN luminosity on redshift, we construct discrete redshift bins with samples that are sensitive down to a common limiting luminosity. We start by choosing a uniform limiting \wisetwo~flux of \FWTwoLoLim$\,$Jy that approximately aligns with the faintest galaxies at $z=0.3$ (left panel of Figure \ref{fig:LBOLAGN_Z_PLOT}). In each redshift bin samples are defined by assigning a lower luminosity limit such that the limiting flux of \FWTwoLoLim$\,$Jy can detect all sources in that bin (right panel of Figure \ref{fig:LBOLAGN_Z_PLOT}). These bins are constructed out to $z=2$ (where the sample is well-represented).

\EBV~is plotted against \LBolAGN~and \FEdd~in Figure \ref{fig:EBV_LBOLAGN_FEDD_PLOT}. When parameterized by a powerlaw function, the slopes are offset from zero at significances of \LBolAGNEBVSlopeSigOne, \LBolAGNEBVSlopeSigTwo, \LBolAGNEBVSlopeSigThree, and \LBolAGNEBVSlopeSigFour$\sigma$ for the redshift bins in ascending order. The decreasing strength of the positive correlation between \EBV~and \LBolAGN~toward higher redshifts (and generally higher \LBolAGN) is consistent with radiation from luminous AGN sublimating dust grains or exerting outward radiation pressure on them. Since we have limited the analysis to QSOs with detectable broad emission lines, the obscuration levels for our sample are generally lower than the threshold used for selecting obscured AGN in the X-ray samples \citep[\NH$\,>10^{22}$\,\uNH;][]{Beckmann:2009,Burlon:2011,Kawamuro:2016}.

On the other hand, negative correlations between \EBV~and \FEdd~are observed (in the same redshift bins the powerlaw slopes are offset from zero at significances of \FEddEBVSlopeSigOne, \FEddEBVSlopeSigTwo, \FEddEBVSlopeSigThree, and \FEddEBVSlopeSigFour$\sigma$). These negative correlations (strongest for the higher redshift bins with generally higher values of \LBolAGN) are consistent with the implication drawn from the X-ray AGN sample of \citet{Ricci:2017c} that the obscuring medium of AGN is within the SMBH gravitational sphere of influence and most efficiently removed from less massive SMBHs accreting close to the Eddington limit. Furthermore, the Eddington ratios in our sample are above the dust effective Eddington limit \citep[\FEdd$\,\sim10^{-2}$;][]{Fabian:2009} and consistent with the obscuring material being removed by radiation pressure.

\section{Conclusions}
\label{sec:conc}

We present a catalog of redshifts and physical properties for the AGN (bolometric luminosities and extinctions) and host galaxies (stellar masses and \SFRs) for \FullSZ~\wise-selected AGN candidates from \citet{Assef:2018}. If spectroscopic redshifts from the \sdss~are available, they are used (\ZSpecSZ~sources). Otherwise, we use photometric redshifts derived from our SED models (\ZPhotSZ~sources). The physical properties are measured from the SED models that separate the AGN component from the host galaxy emission and compute intrinsic bolometric luminosities and dust extinctions. From the isolated host galaxy component we compute stellar masses and \SFRs~by accounting for stellar emission plus dust absorption and re-emission. 

In this paper we detail how the sample is developed and describe the AGN and host galaxy physical properties. We then use the sample to identify purely star-forming galaxy contaminants admitted by the \wise~AGN selection criteria. Finally, we use the sample to examine the connections between host galaxy properties, SMBH growth, and nuclear obscuration. Our conclusions are as follows:

\begin{itemize}

\item Using the subset with spectroscopic redshifts, we find the photometric redshift accuracy to be \SigNMADSymb$\,=\,$\SigNMAD, with an outlier fraction of \OLFSymb$\,=\,$\OLF. The outliers have systematically under-estimated photometric redshifts due to the prior and are primarily associated with sources having blue \rwtwo~colors and large AGN fractions at photometric redshifts of \ZPhot$\,\lesssim1$.

\item While the subset of \wise~AGN with spectroscopic redshifts is dominated by relatively unobscured QSOs (median \EBV$\,=\,$\EBVMedianSpec), the subset with only photometric redshifts in our catalog have generally larger AGN obscuration (median \EBV$\,=\,$\EBVMedianCorrPhot). This photometric redshift subset has a median AGN bolometric luminosity of \LBolAGN$\,=\,$\LBolAGNMedianCorrPhot\,\uLum~and therefore contains many previously unidentified luminous and obscured AGN. A fraction (\QSOHiEBVPerc$\%$) of spectroscopic QSOs have obscuration values above the threshold broadly separating \typeI~and \typeII~AGN (\NH$\,>10^{22}$\,\uNH) and may represent examples of obscured yet intrinsically luminous AGN with strong broad emission lines.

\item Based on spectroscopic AGN signatures (broad emission lines or narrow emission line diagnostics), \GAMAReliableFrac\%~of the \wise-selected AGN candidates are true AGN. This probability is a strong function of the AGN fractions measured from our SED models, with AGN fractions $>$\,0.5 corresponding to an average probability of \GAMAHiAHatAGNPerc$\%$ for hosting an AGN.

\item Contaminants have relatively blue \wiseone$-$\wisetwo~colors (median value of \wiseone$-$\wisetwo$\,=\,$\WOneTwoMedianLoAHat, compared to a median value of \wiseone$-$\wisetwo$\,=\,$\WOneTwoMedianHiAHat~for likely AGN). For contaminants, purely star-forming SED models can yield excellent fits in the redshift range \z$\,\sim0.2-0.3$ where the majority of candidates with low AGN fractions are found. The purely star-forming SED models suggest that the \wiseone$-$\wisetwo~colors are achieved by a combination of stellar attenuation and dust emission, and these potential contaminants are biased toward high specific \SFRs.
 
\item Moving toward lower redshifts, the specific \SFRs~of \wise~AGN host galaxies show an increasing negative offset relative to the prediction for normal star forming galaxies. These offsets are most significant at the lowest redshifts ($\sim\,$50\,\%) and suggest evolution of the host galaxy population toward a passive sequence due to reduced cold gas supplies at later epochs. This evolution may also suggest a temporal delay between the onset of star formation and luminous AGN triggering due the time necessary for cold gas to travel from global scales (where the majority of star formation occurs) to nuclear scales for SMBH accretion.

\item AGN extinctions show a marginal positive correlation with AGN bolometric luminosities ($<$\,3$\sigma$ significance) that becomes weaker for more luminous AGN. However, AGN extinctions show negative correlations with Eddington ratios that are strongest among the most luminous AGN ($>$\,6$\sigma$ significance). This result is consistent with the `receding torus' model being regulated by radiation pressure and also with the obscuring material being within the SMBH gravitational sphere of influence.

\end{itemize}

\acknowledgments
{We thank an anonymous reviewer for detailed and helpful comments that have greatly improved the quality of the manuscript. This work is supported by the NASA Astrophysics Data Analysis Program 18-ADAP18-0138. RJA was supported by FONDECYT grant number 1191124.}

\facilities{\galex, Sloan, \emph{Gaia}, PS1, CTIO:2MASS, FLWO:2MASS, \wise, \emph{Herschel}}

\software{\LRTtitle~\citep{Assef:2008,Assef2010}, \astropy~\citep{astropy:2013, astropy:2018}.}

\end{document}